\pdfoutput=1

\documentclass[cits]{PoS}

\usepackage{lscape}

\title{Partially Screened Gap - general approach and observational 
    consequences}

\ShortTitle{Partially Screened Gap}

\author{\speaker{Andrzej Szary}, George I. Melikidze\thanks{
            E.Kharadze Georgian National Astrophysical Observatory, Georgia}~ 
            and Janusz Gil\\
        J.Kepler Institute of Astronomy, University of Zielona G\'{o}ra, 
            Zielona G\'{o}ra, Poland\\
        E-mail: \email{aszary@astro.ia.uz.zgora.pl},
            \email{gogi@astro.ia.uz.zgora.pl}, 
            \email{jag@astro.ia.uz.zgora.pl}}

\abstract{
        Observations of the thermal X-ray emission from radio pulsars implicate
        that the size of hot spots is much smaller then the size of the polar
        cap that follows from the purely dipolar geometry of pulsar magnetic 
        field. 
        Most plausible explanation of this phenomena is an assumption that
        the magnetic field at the stellar surface differs essentially from the 
        purely dipolar field. 
        We can determine magnetic field at the surface by the conservation of 
        the magnetic flux through the area bounded by open magnetic field
        lines. 
        Then the value of the surface magnetic field can be estimated as of 
        the order of $10^{14}$ G. On the other hand observations show 
        that the temperature of the hot spot is about a few million Kelvins.
        Based on these observations the Partially Screened Gap (PSG) model was
        proposed which assumes that the temperature of the actual polar cap
        (hot spot) equals to the so called critical temperature.
        
        We discuss correlation between the temperature and corresponding area
        of the thermal X-ray emission for a number of pulsars.
        The results of our analysis show that the PSG model is suitable
        to explain both cases: when the hot spot is smaller and larger then 
        conventional polar cap. We argue that in the second case structure 
        and curvature of field lines allow pair creation in the closed 
        field lines region thus the secondary particles can heat the stellar
        surface outside the actual polar cap.

        We have found that the Curvature Radiation (CR) plays dominant role in
        avalanche pair production in the PSG. We studied dependence of the
        PSG parameters on the pulsar period, the magnetic field strength and 
        the curvature of field lines. 
}

\FullConference{High Time Resolution Astrophysics IV - The Era of Extremely 
                Large Telescopes - HTRA-IV,\\
		May 5-7, 2010\\
		Agios Nikolaos, Crete, Greece}

\begin{document}

    \section{Introduction}
        The Standard model of radio pulsars assumes that there exists the 
        Inner Acceleration Region (IAR) above the polar cap where the electric
        field has a component along the opened magnetic field lines. In this
        region particles (electrons and positrons) are accelerated in both
        directions: outward and toward the stellar surface. Consequently,
        outflowing particles are responsible for generation of the
        magnetospheric emission (radio and high-frequency) while the
        backflowing particles heat the surface and provide required energy
        for the thermal emission. In such scenario analysis of X-ray radiation
        is an excellent method to get insight into the most intriguing region
        of the neutron star.\\

    \subsection{Observations}
        About 30 years ago the first X-ray telescope called \verb+Einstein+ 
        was put into space thus opening the possibility for direct 
        investigation of thermal emission from isolated neutron stars.
        A significant contribution to this study was provided by \verb+ROSAT+ 
        in 1990's. Currently operating observatories such as \verb+Chandra+ and
        \verb+XMM-Newton+ have greatly increased the quality and availability
        of observations of thermal radiation from neutron star surfaces.
        
        The X-ray radiation from an isolated neutron star in general can 
        consist of two distinguishable components: the nonthermal emission and 
        the thermal emission. The nonthermal  component is usually described
        by a power-law spectral model and attributed to radiation produced in
        the pulsar magnetosphere while the thermal emission can originate
        either from the entire surface of cooling neutron star or the small
        hot spots  around the magnetic poles on stellar surface 
        (polar caps and adjacent areas).

        Thermal X-ray emission seems to be a quite common feature of radio 
        pulsars. The black body fit allows us to obtain directly the 
        temperature ($T_s$) of the hot spot. Using the distance ($D$) to 
        the pulsar and the luminosity of thermal emission ($L_{\rm bol}$) we 
        can estimate the area ($A_{\rm BB}$) of the hot spot. In the most 
        cases $A_{\rm BB}$ differs from the conventional polar cap area 
        $A_{\rm pc}\approx 6.2 \times 10^{4} P^{-1} $ m$^2$, where $P$ is 
        pulsar period. We use parameter  $b = A_{\rm pc} / A_{\rm BB}$ to 
        describe the difference between $A_{\rm pc}$ and $A_{\rm BB}$.

        \subsection{The case with $b < 1$ \label{sub_b}}
         
        In the most cases observed hot spot area ($A_{\rm BB}$) is larger then 
        the conventional polar cap area. We can distinguish two 
        types of pulsars in this group, with $b\ll1$ and $b \lesssim 1$.
        
        The first type is associated with observations of thermal emission from
        the entire  stellar surface and can be used to test cooling models.
        Although we have to remember that for young pulsars ($\tau = 1 $ kyr)
        the nonthermal component dominates, making it impossible to measure 
        accurately the thermal flux. 
        As a pulsar becomes older, its nonthermal luminosity decreases faster 
        then the thermal  luminosity up to the end of the neutrino-cooling era
        ($\tau \sim 1$ Myr). Thus, the thermal radiation from the entire
        stellar surface can dominate at soft X-ray energies for middle-age
        pulsars ($\tau \sim 100$  kyr) and some younger pulsars 
        ($\tau \sim 10$ kyr) \cite{2007_Zavlin_b}.

\begin{figure}[!ht]
                \centering
	        \includegraphics[height=7cm, width=12cm]{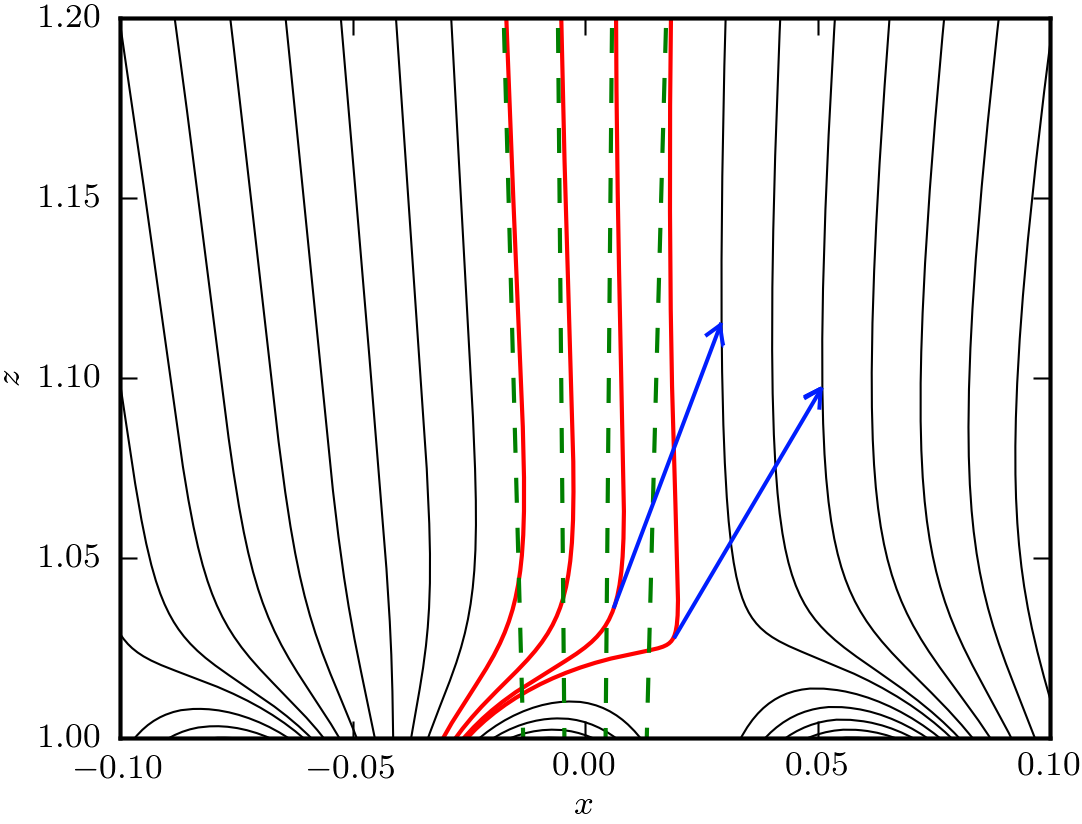}
                \caption{Cartoon of the magnetic field lines in the polar cap
                region. Red lines are open field lines and green dashed lines
                correspond to the dipole field. The blue arrows show direction
                of the curvature photons emission.}
                \label{non_dipolar_heat}
            \end{figure}

        Some observations show that the hot spot area is larger then the 
        conventional polar cap area but still significantly less then the 
        area of the star. In this case the radiation comes from the surface
        adjacent to polar cap. Therefore, it can not be explained by cooling
        of the surface and some addition heating mechanism is needed. The model
        of such heating is based on the assumption that the pulsar magnetic 
        field near the stellar surface differs significantly from the pure
        dipole one. The calculations show that it is natural to obtain such 
        geometry of magnetic field lines that allows the pair creation in the
        closed field lines region.  The pairs move along closed magnetic field 
        lines and heat the surface beyond the polar cap on the opposite side of
        the star (Fig. \ref{non_dipolar_heat}).
        In such scenario heating energy is generated in IAR (outward particles)
        hence the luminosity should be same order as the nonthermal one but it
        is hard to predict which of them would prevail in X-ray flux of an old
        neutron star. However, it cannot be ruled out that the thermal one may 
        be dominant as suggested by some observations.
        In most cases area of a such heated surface may be larger (but not 
        necessarily) then the conventional polar cap area. That makes the
        estimation of parameters of black-body radiation even harder. More
        detailed description of this phenomenon can be found in 
        \cite{2007_Melikidze}.

        \subsection{The case with $b > 1$}

        In some cases observed hot spot area ($A_{\rm BB}$) is less then the
        conventional polar cap area ($b > 1$). The model mentioned in 
        \ref{sub_b} can be used in order to explain this phenomenon, because
        the size of modified polar cap is much smaller then size of 
        conventional polar cap (see Fig. \ref{non_dipolar_heat}).     
        The surface magnetic filed can be estimated by the magnetic flux 
        conservation law as $b = A_{\rm pc} / A_{\rm BB}$ = $B_s / B_d$. Where 
        $B_d = 2.02 \times 10^{12} (P\dot{P}_{-15})^{0.5}$, $P$ is the pulsar
        period in seconds and $\dot{P}_{-15} = \dot{P}/10^{-15}$ is the period
        derivative. 

\begin{figure}[!ht]
                \centering
                \includegraphics[height=6cm, width=7.4cm]{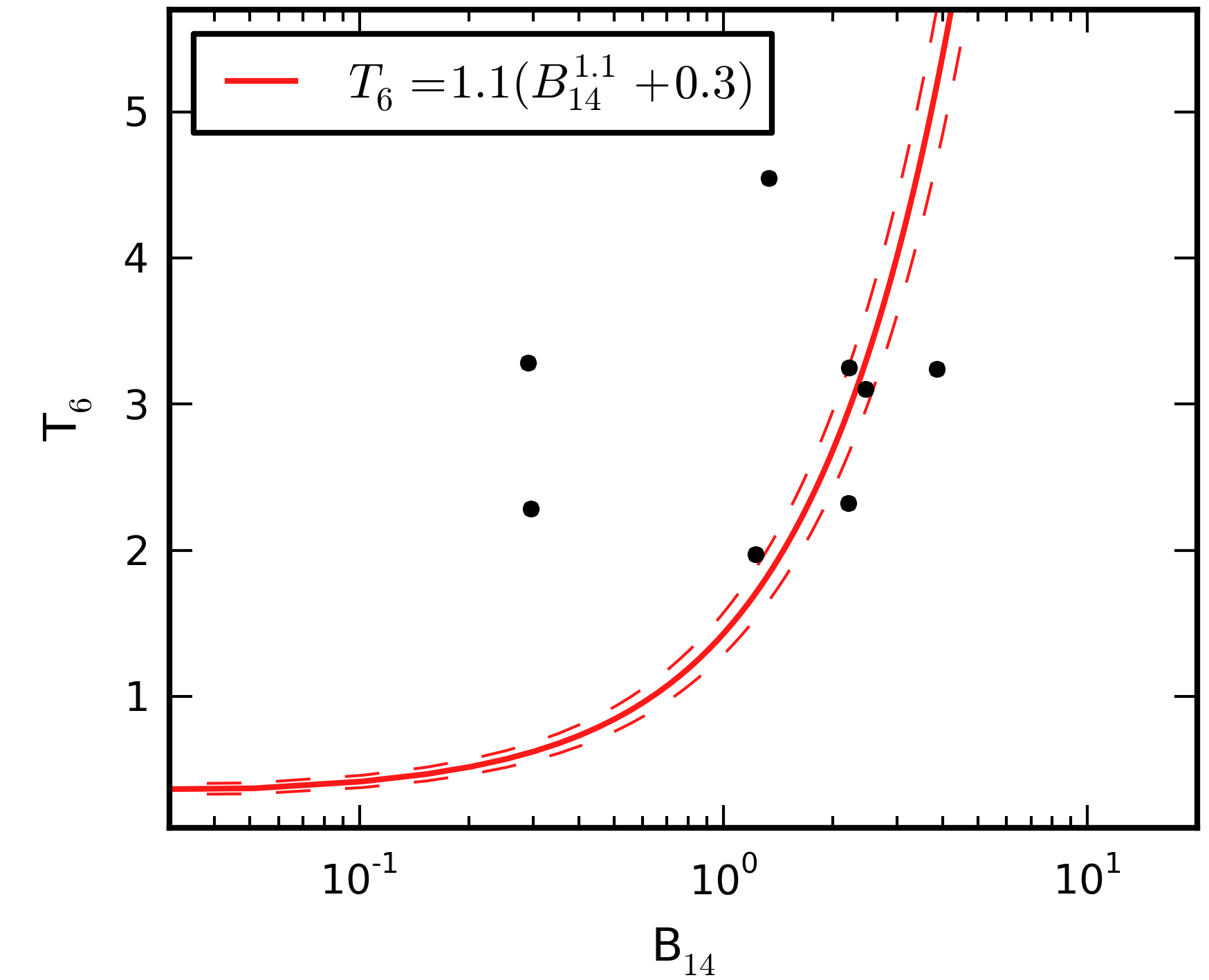}
                \hspace{0.1cm}
                \includegraphics[height=6cm, width=7.4cm]{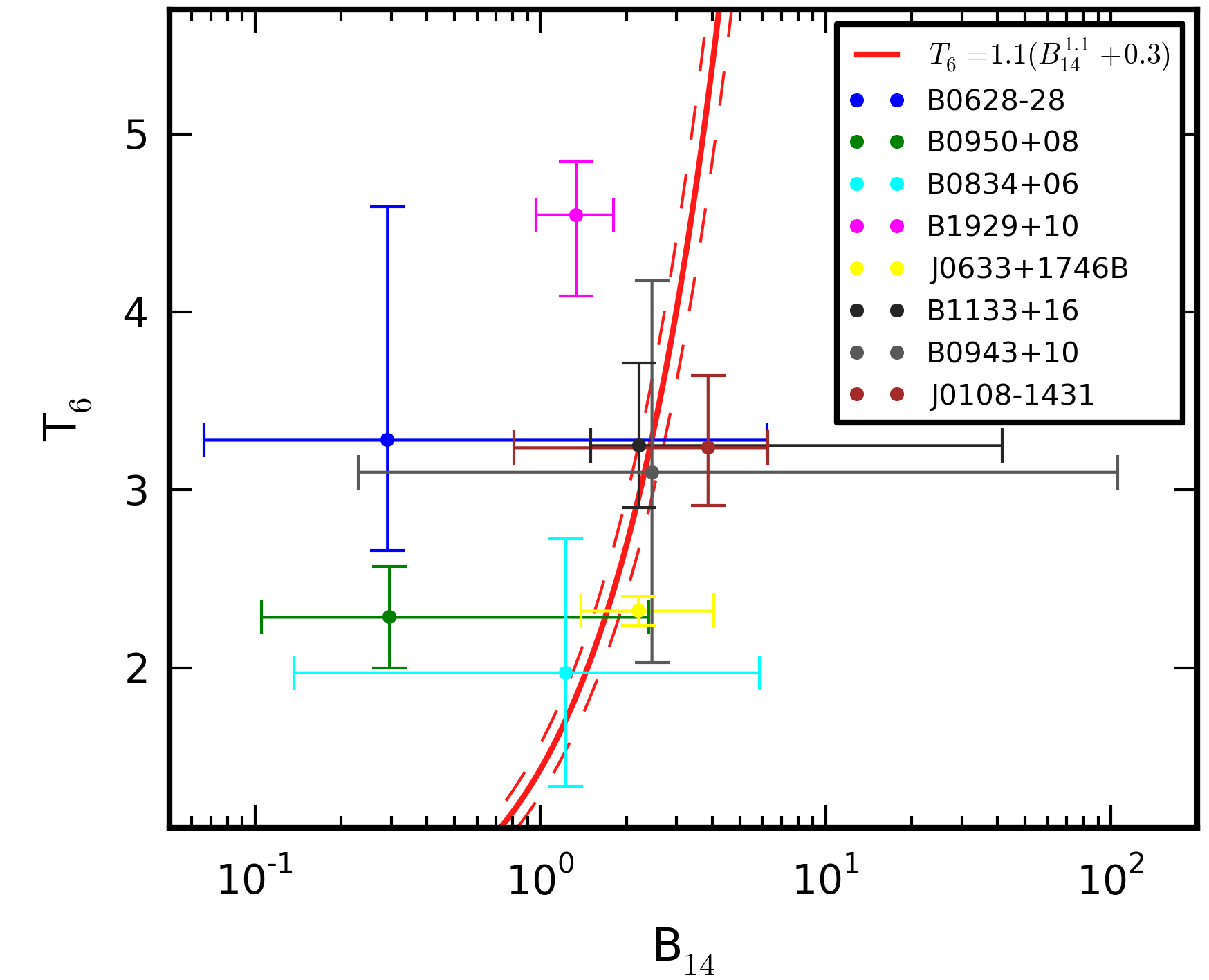}
                \caption{Diagram of the surface temperature 
                    ($T_6 = T_s / 10^6$) vs. the surface magnetic field 
                    ($B_{14}=B_s/10^{14}$). The red line is the critical 
                    temperature ($T_{\rm crit}$) evaluated from 
                    \cite{2008_Medin} and dashed lines corresponds to 
                    uncertainties in calculations.
                    Left panel
                    includes all pulsars with $b>1$ while right panel is zoom
                    of the upper right part of the graph. Error bars
                    corresponds to $1\sigma$.}
                \label{medin_lai}
            \end{figure}

        Medin \& Lai \cite{2008_Medin} calculated the condition for the 
        formation of a vacuum gap above neutron star surface. In Fig. 
        \ref{medin_lai} we present positions of pulsars with derived surface 
        temperature ($T_s$) and hot spot area ($A_{\rm BB}$) on the $B_s-T_s$
        diagram where $B_s$ is estimated as $B_s = b B_d$. Red line represents
        dependence of $T_{\rm crit}$ on $B_s$. We can see that in most cases
        the pulsars positions follow with $B_s-T_{\rm crit}$ theoretical
        curve. Few cases which do not coincide with the theoretical curve 
        can be explained by heating the surface outside the polar cap 
        (see \ref{sub_b}).
        
        According to our model the actual surface temperature equals 
        to the critical value ($T_s \sim T_{\rm crit}$) which leads to the 
        formation of Partially Screened Gap (PSG) above the polar caps of
        neutron star \cite{2003_Gil}. Hot spot parameters derived from  X-ray 
        observations of isolated neutron stars are presented in Table 
        \ref{obs_table}.

    \section{The Model}

        The PSG model assumes existence of heavy (Fe$^{56}$) ions with density
        near
        but still below corotational charge density ($\rho_{\rm GJ}$), thus 
        the actual charge density causes partial screening of the potential 
        drop just above the polar cap. The degree of shielding can be 
        described by parameter $\eta = 1 - \rho_i/\rho_{\rm GJ}$ where $\rho_i$
        is the charge density of heavy ions in the gap.
        The thermal ejection of ions from surface causes partial screening
        of the acceleration potential drop
            \begin{equation}
                \Delta V = \eta \Delta V_{max} = 2 \pi \eta B_s h^2 /c P, 
                \label{d_v1}
            \end{equation}
        where $\Delta V_{max}$ is the potential drop in vacuum gap, $h$ is the 
        gap height, $B_s = b B_d$ 
        surface magnetic field (applicable only if $b>1$). Using calculations
        of Medin \& Lai \cite{2008_Medin} we can express the dependence of the
        critical temperature on pulsar parameters as
            \begin{equation}
                T_{\rm crit} = 1.6 \times 10^4
                    \left( \left( \left(P \dot{P}_{-15}\right)^{0.5} b\right)
                    ^{1.1} + 17.7\right)
                \label{t_s}
            \end{equation}
        or $T_{\rm crit} = 1.1 \times 10^6 \left(  B_{14}^{1.1} + 0.3\right)$, 
        where $B_{14}=B_s/10^{14}$.
        
        The actual potential drop $\Delta V$ should be thermostatically 
        regulated and there should be established a quasi-equilibrium state,
        in which heating due to electron/positron bombardment is balanced by
        cooling due to thermal radiation (see \cite{2003_Gil} for more 
        details). The necessary condition for formation of this 
        quasi-equilibrium state is
            \begin{equation}
                \sigma T_s^4 = \eta e \Delta  V   c n_{\rm GJ},
                \label{heating_condition}
            \end{equation}
        where $\sigma$ is the Stefan-Boltzmann constant, $e$ - the electron 
        charge, $n_{\rm GJ} = \rho_{\rm GJ}/e = 1.4 \times 10^{11} b 
        \dot{P}_{-15}^{0.5} P^{-0.5}$ is the corotational number density.
        Using equations (\ref{t_s}) and (\ref{heating_condition}) we can 
        express the acceleration potential drop as
            \begin{equation}
                \Delta V = 7.3 \times 10^5 \eta^{-1} B_{14}^{-1} 
                    \left( B_{14}^{1.1} + 0.3\right)^4 P,
                \label{d_v2}
            \end{equation}
        and finally using equations (\ref{d_v1}) and (\ref{d_v2}) we can
        estimate the gap height  in PSG model as
            \begin{equation}
                h \eta = 6 B_{14}^{-1} 
                    \left( B_{14}^{1.1} + 0.3\right)^2 P 
                \label{gap_height_eqs}
            \end{equation}

        As we see both $\Delta V$ and $h$ depend on shielding factor $\eta$ 
        which actually depends on the details of the avalanche pair production
        in the gap. First we need to determine which process, Curvature 
        Radiation (CR) or Inverse Compton Scattering (ICS), is responsible 
        for the pair production. We will need following parameters: 
        $l_{\rm acc}$ - the distance which a particle should pass to gain 
        Lorentz factor equal to $\gamma_{\rm acc}$, $l_e$ - the mean length an 
        electron (or positron) travels before a gamma-photon is emitted,
        and 
        $l_{\rm ph}$ - the mean free path of gamma-photon before being
        absorbed by the magnetic field.

            \subsection{Acceleration path}
        
        In the frame of PSG model, using formalism described in 
        \cite{1975_Ruderman}, we estimate the component of electric field 
        along the magnetic field line in the gap as
            \begin{equation}
                E \approx \eta \frac{4 \pi B_s}{c P} (h-z)
            \end{equation}
        which vanishes at the top $z=h$.
        The Lorentz factor of particles after distance $l_{\rm acc}$ can be 
        calculated as follows

            \begin{equation}
                \gamma_{\rm acc} = \frac{e}{m_e c^2} \int_{z_1}^{z_2} E dz 
                    \approx \eta \frac{4 \pi B_s e}{m_e c^3 P} (z_2 -z_1) (h - 
                    \frac{z_1+z_2}{2})
            \end{equation}
        where $m_e$ is mass of an electron and $z_2 - z_1 = l_{\rm acc}$. Then
        we can approximate $z_1 + z_2 \approx 2z$ and assume $z \approx h/2$, 
        thus
            \begin{equation}
                l_{\rm acc} = \gamma_{\rm acc} \frac{m_e c^3 P} 
                    {2 \pi h \eta B_s e}. 
                \label{acceleration_eqs}
            \end{equation}

            \subsection{Electron or positron mean free paths}

        The mean free path of electron or positron ($l_e$) can be defined as 
        the mean length that a particle passes until a gamma-photon is 
        emitted. In the case of CR electron mean free path can be estimated as
        a distance that particle with Lorentz factor $\gamma$ travels
        during the time which is necessary to emit curvature photon 
        (see \cite{1997_Zhang})
            \begin{equation}
                l_{e,cr} \sim c \left ( \frac{P_{cr}}{E_{\gamma,cr}} 
                    \right)^{-1} 
                    = \frac{9}{4} \frac{\hbar \Re c}{\gamma e^2},
                \label{cr_free_path}
            \end{equation}
        where $P_{cr} = 2 \gamma^4 e^2 c / 3 \Re^2$ is the power of the CR,
        $E_{\gamma, cr} =  3\hbar \gamma^3 c / 2\Re$ is the 
        characteristic photon energy, $\Re$ - the curvature radius of magnetic 
        field lines.

        For the ICS process, calculation of the electron mean free path
        $l_{e,ics}$, is not as simple as that of the CR process. Although
        we can define $l_{e,ics}$ in a same way as we defined $l_{e,cr}$ but
        it is difficult to estimate a characteristic frequency of emitted
        photons. We have to take into account photons of various frequencies
        with various incident angles. An estimation of the mean free path 
        of electron (or positron) to produce a photon is \cite{1985_Xia}
            \begin{equation}
                l_{e,ics} \sim \left [ \int_{\mu_0}^{\mu_1} \int_0^{\infty}
                    {\sigma'(\epsilon,\mu_i) 
                    ( 1-\beta \mu_i) n_{\rm ph}(\epsilon) d \epsilon d \mu_i}
                    \right ]^{-1}
            \end{equation}
        where $\epsilon$ is the incident photon energy in units of $m_e c^2$, 
        $\mu_i = \cos{\psi_i}$ is the cosine of the
        photon incident angle, $\beta = v/c$ is the velocity in terms of speed
        of light,

            \begin{equation}
                n_{\rm ph} = \frac{4 \pi}{\lambda_c^3} 
                \frac{\epsilon^2}{\exp{(\epsilon/\theta)}-1} d \epsilon
            \end{equation}
        represents the photon number density distribution of a semi-isotropic
        blackbody radiation, $\theta = k T_s/m_e c^2$, $k$ is the\
        Boltzmann constant, and 
        $\lambda_c = h/m_e c = 2.424 \times 10^{-10}$ cm is the electron
        Compton wavelength. Here $\sigma '$ is the cross section of scattering 
        in the particle rest frame. In the Thomson regime cross section of 
        scattering in the particle rest frame can be written as
        \cite{1989_Dermer}
            \begin{equation}
                \sigma ' = \frac{\sigma_T}{2} \left [ \frac{u^2}{(u+1)^2} +
                    \frac{u^2}{(u-1)^2 + a^2} \right ] ,
                \label{ics_cross}
            \end{equation}
        where $\sigma_T$ is the Thomson cross section, 
        $u = \epsilon ' /\epsilon_B$, $a= \frac{2}{3}\alpha_f \epsilon_B$, 
        $\epsilon ' = \gamma \epsilon (1-\beta \mu_i)$ is the incident photon 
        energy in the particle rest frame in units of $m_e c^2$, 
        $\epsilon_B = B_s / B_q$ is the electron cyclotron resonance energy
        in units of $m_e c^2$, $B_q=m_e^2c^3/e \hbar = 4.413 \times 10^{13}$
        G, and $\alpha_f=e^2/\hbar c$ is the fine-structure constant. 
        
        Equation (\ref{ics_cross}) however does not include the quantum 
        relativistic effects
        of a strong magnetic field ($B_s > 0.1 B_q$), therefore, it can be used
        only at altitudes where magnetic field is relatively weaker. 
        For strong magnetic fields we can use an approximation proposed in
        \cite{2000_Gonthier} and then an approximate averaged 
        (polarization-independent) cross section can be written as follows
            
            \begin{equation}
                \sigma ' =  \frac{3 \sigma_T}{8} \frac{\epsilon \epsilon'^2 
                    (1 + \mu_i^2)}{2 \epsilon - \epsilon'} \left [ 
                    \frac{1}{(\epsilon - \beta_q)^2} + \frac{1}
                    {(\epsilon + \beta_q)^2} \right ],
                \label{cross_gonthier}
            \end{equation}
        where $\beta_q = B_s/B_q$. Equation (\ref{cross_gonthier}) represents
        the exact cross section when the particle after scattering falls on the
        zero Landau state (see \cite{2000_Gonthier} for more accurate results).

        We should expect two modes of ICS process, resonant and non-resonant.
        The resonant ICS takes place if the photon frequency in particle rest
        frame equals to the cyclotron electron frequency.
        Non-resonant ICS includes all scattering processes of photons with 
        frequencies around the maximum of the thermal spectrum.
        At resonance frequency equation 
        (\ref{cross_gonthier}) has a singularity, so it could be used only for
        non-resonant case. For resonant case, one should use approach proposed
        in \cite{2010_Medin}, thus the cross section in the particle rest frame
        is
            \begin{equation}
                \sigma_{res} ' \simeq 2 \pi^2 \frac{e^2 \hbar}{m_e c} \delta 
                    (\epsilon ' - \epsilon_b)
                \label{res_cross_a}
            \end{equation}
        Equation (\ref{res_cross_a}) can be also used for strong magnetic 
        fields ($\beta_q > 0.1$), since the resonant condition
        $\epsilon' = \epsilon_b$ holds regardless of field strength. Although
        one have to remember that this equation represents the upper limit for
        cases with very high magnetic fields (see \cite{2000_Gonthier} for
        more details). The particle mean free path above a polar cap for the 
        resonant ICS process is
            \begin{equation}
                l_{e,rics} \sim \left [ \int_{\mu_0}^{\mu_1} \int_0^{\infty}
                    {\sigma_{res}' 
                    ( 1-\beta \mu_i) n_{\rm ph}(\epsilon) d \epsilon d \mu_i}
                    \right ]^{-1} = \left [\frac{2 \pi e^2 \hbar}{m_e c \gamma}
                    \int_{\mu_0}^{\mu_1} n_{\rm ph} \left(
                    \frac{\epsilon_b}{1-\beta \mu_i} \right) d \mu_i \right ]^{-1}
                \label{ics_free_path}
            \end{equation}
        For the altitudes of the same order as the polar cap size we can use
        $\mu_0=1$, $\mu_1=0$ as incident angle limits for outflowing particles,
        and $\mu_0=0$, $\mu_1=-1$ as incident angle limits for backflowing 
        particles.

            \subsection{Photon mean free path}

        A photon with energy $E_{\gamma}  > 2 m_e c^2$ and propagating with
        nonzero angle $\psi$ with respect to an external magnetic field can
        be absorbed by the field and as a result electron-positron pair is
        created. The optical depth of a such process can be defined as 
        \cite{2010_Medin}
            \begin{equation}
                \tau = s_{\rm ph}R_{\|, \perp},
            \end{equation}
        where $s_{\rm ph}$ is a distance traveled by an photon,
        $R_{\|, \perp} = R'_{\|, \perp} \sin{\psi}$ is the attenuation
        coefficient for the $\|$ or $\perp$ polarized photons, and $R'$ is the 
        attenuation coefficient in the ''perpendicular'' frame 
        (the frame where the photon propagates perpendicular to the local
        magnetic field). 

        The total attenuation coefficient for pair production is given by
        $R' = \sum_{jk} R'_{j,k}$, where $R'_{j,k}$ is the attenuation
        coefficient for process (channel) in which the photon produces an 
        electron in Landau level $j$ and positron in Landau level $k$, and the
        sum is taken over all possible states for the electron-positron pair. 
        Since pair production is symmetric with respect to the electron and 
        positron, $R'_{kj} = R'_{kj}$ for simplicity we will use $R'_{jk}$ to
        represent the combined probability of creating the pair in either 
        $(jk)$ or $(kj)$ state. For a given channel $(jk)$, the threshold 
        condition for pair production is:
            \begin{equation}
                E_{\gamma} ' > E_j' + E_k',
                \label{energy}
            \end{equation}
        where $E_{\gamma} ' = E_{\gamma} \sin{\psi}$ is the photon energy in
        the perpendicular frame and $E_n' = m_e c^2 \sqrt{1 + 2 \beta_q n}$
        is the minimum energy of an electron/positron in Landau Level $n$.
        In dimensionless form this condition can be written as
            \begin{equation}
                x = \frac{E_{\gamma} '}{2 m c^2} = \frac{E_{\gamma}}
                    {2 m_e c^2} 
                    \sin{\psi} > \frac{1}{2} \left [ \sqrt{1 + 2 \beta_q j}
                    + \sqrt{1 + 2 \beta_q k}\right ]
            \end{equation}
        The first nonzero attenuation coefficients for both polarizations 
        are given in \cite{1983_Daugherty}:
            \begin{equation}
                R'_{\|,00} = \frac{1}{2 a_0} \frac{\beta_q}{x^2 \sqrt{x^2 - 1}}
                    e ^{-2x^2 / \beta_q}, \hspace{1cm} x > x_1 = 1;
            \end{equation}
        
            \begin{equation}
                R'_{\perp,01} = 2 \times \frac{1}{2 a_0} \frac{\beta_q}{2x^2}
                    \frac{2 x ^2 - \beta_q }
                    {\sqrt{x^2 - 1 - \beta_q + \frac{\beta_q^2}{4x^2}}}
                    e ^{-2x^2 / \beta_q}, \hspace{1cm} 
                    x > x_2 = \left( 1+ \sqrt{1 + 2 \beta_q} \right) / 2;
            \end{equation}
        where $a_0$ is Bohr radius (let us note that $R'_{\perp,00} = 0$ and 
        higher orders of attenuation coefficients should be used if 
        $x > x_3 = \left( 1+ \sqrt{1 + 4 \beta_q} \right) / 2$).

        The optical depth is defined as:
            \begin{equation}
                \tau = \int_0^{s_{\rm ph}} R(s) ds = \int_0^{s_{\rm ph}}
                    R'(s) \sin{\psi} ds
                \label{tau_in}
            \end{equation}
        We can assume $\psi \ll 1$, because all high energy photons ($x>1$) 
        will produce pairs much earlier then $\psi$ reaches value near unity.
        In this limit $\sin{\psi} \simeq s_{\rm ph} / \Re$ so relation between 
        $x$ and $s_{\rm ph}$ can be expressed by
            \begin{equation}
                x \simeq \frac{s_{\rm ph}}{\Re} \frac{E_{\gamma}}{2 m_e c^2}.
            \end{equation}
        The equation (\ref{tau_in}) can be rewritten as

            \begin{displaymath}
                \tau = \tau_{1} + \tau_{\|,2} + \tau_{\perp,2} + ...
            \end{displaymath}
            \begin{equation}
                \tau_1 = \int_{s_1}^{s_2} R_{\|,00} ds, \hspace{0.5cm}
                    \tau_{\|,2} = \int_{s_2}^{s_3} \left 
                    ( R_{\|,00} +  R_{\|,01}  \right ) ds, \hspace{0.5cm}
                    \tau_{\perp,2} = \int_{s_2}^{s_3}  R_{\perp,01}  ds 
            \end{equation}
        where $s_1$ and $s_2$ are distances which the photon should pass in 
        order to have energy $x_1$ and $x_2$ respectively (in perpendicular 
        frame of reference). Let us note that $s_1$, $s_2$, and $s_3$ are 
        of the same order and if $s < s_1$ attenuation coefficient is zero.
        As shown in \cite{2010_Medin} for strong magnetic fields 
        ($\beta_q \gtrsim 0.1$) $\tau_{1}$, $\tau_{\|,2}$, and $\tau_{\perp,2}$
        are much larger then one.
        Therefore, the pair production process happens according two scenarios.
        If $\beta_q \gtrsim 0.1$ photons produce pairs almost immediately upon 
        reaching the first threshold, so the created pairs will be in low 
        Landau levels ($n \lesssim 2$). If $\beta_q \lesssim 0.1$ photons
        will travel longer distances to be absorbed and created pairs will be
        in higher Landau levels.

        Thus, for strong magnetic fields ($\beta_q \gtrsim 0.1$) electron mean 
        free path can be calculated as
            \begin{equation}
                l_{\rm ph} \sim s_{1} = \Re \frac{2 m_e c^2}{E_{\gamma}},
                \label{photon_path}
            \end{equation}
        and for relatively weak magnetic fields ($\beta_q \lesssim 0.1$) 
        we can use asymptotic approximation derived by Erber \cite{1966_Erber}
            \begin{equation}
                l_{\rm ph} = \frac{4.4}{(e^2/\hbar c)} 
                    \frac{\hbar}{m_e c}\frac{1}{\beta_q \sin{\psi}} 
                    \exp{\left( \frac{4}{3 x \beta_q }\right)}
                \label{photon_path2}
            \end{equation}

        \subsection{Partially Screened Gap height \label{finding_height_sec}}
            
\begin{figure}[!ht]
            \centering
            \includegraphics[width=7.5cm, height=6.cm]{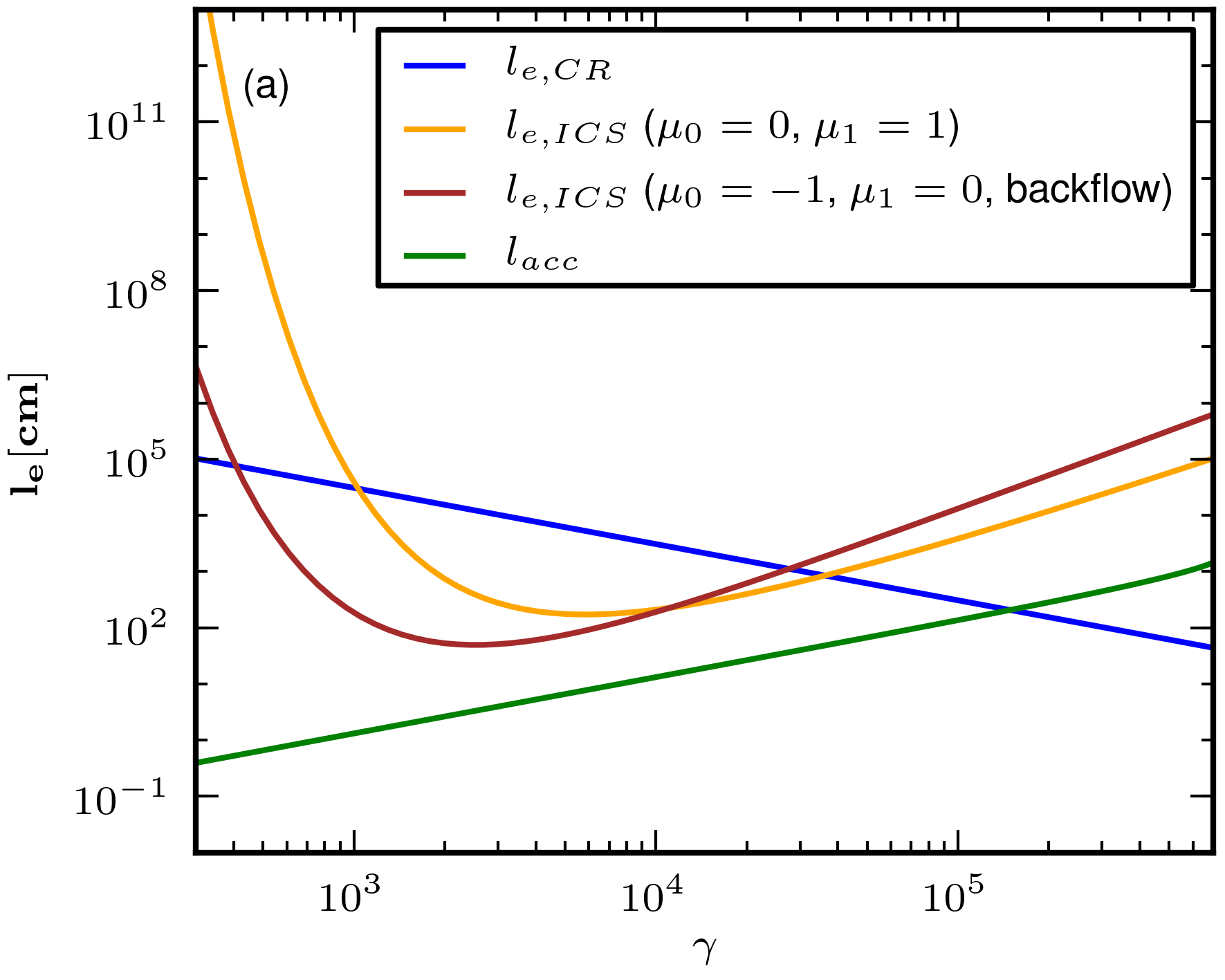}
            \includegraphics[width=7.5cm, height=6.cm]{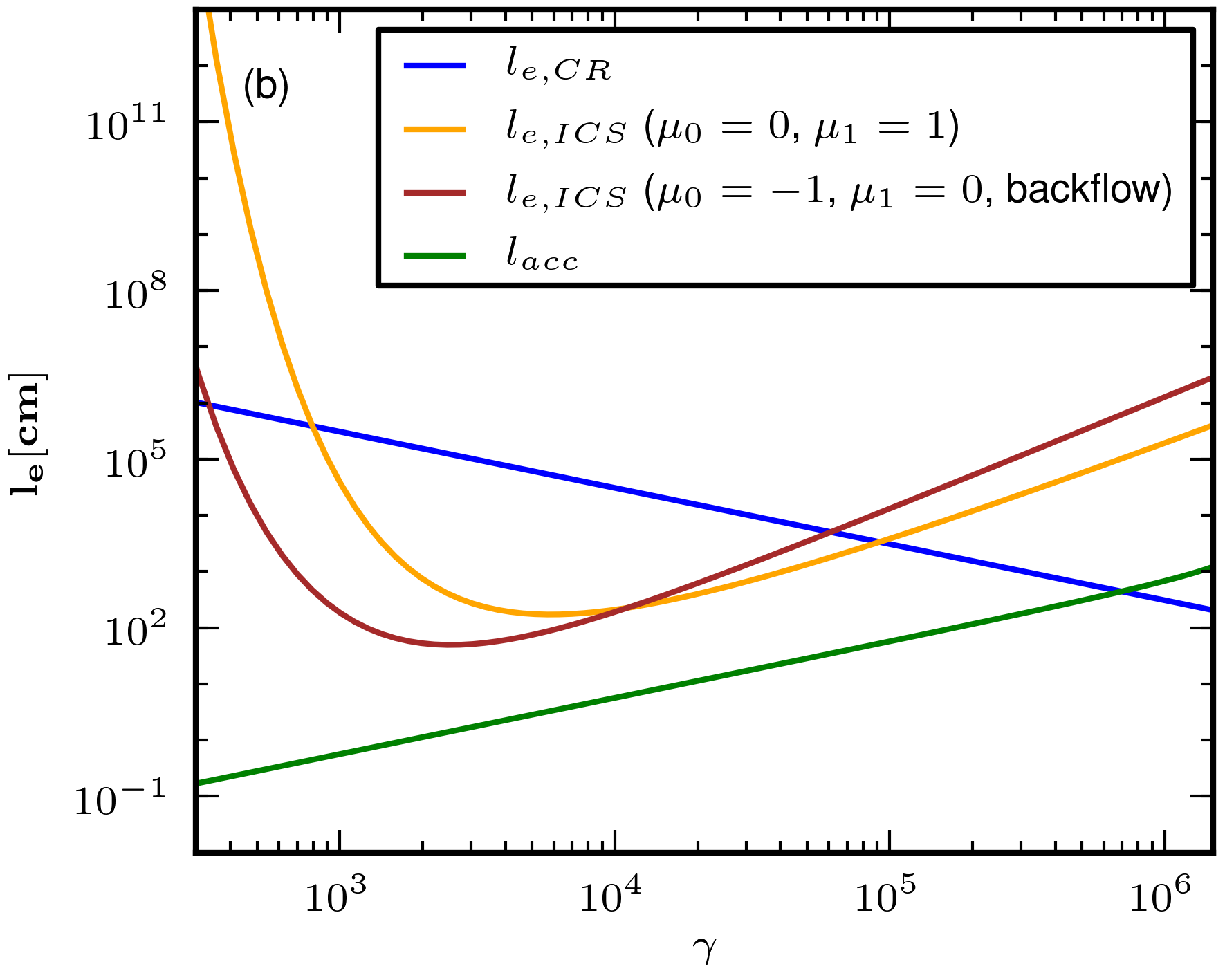}

            \caption{Diagram of the electron/positron mean free path for both 
                CR and ICS processes. The left panel (a) corresponds to 
                calculations for $B_{14}=1.5$, $T_6=1.9$, $\Re_6=0.1$, $P=1$ 
                while the right panel (b) corresponds to calculations for 
                $B_{14}=3.5$, $T_6=4.4$, $\Re_6=1$, $P=1$. Acceleration path 
                was calculated for the same gap parameters ($h=20$ m, 
                $\eta=0.03$) in both panels (see discussion in section 
                \protect \ref{finding_height_sec}).
                Although $l_{e, ics} < l_{e, cr}$ for particles with 
                $\gamma \sim 10^3 - 10^4$ the $\gamma$-photon production by ICS
                process is not effective because for these particle 
                $l_{\rm acc} << l_{e, ics}$. That means that they will be 
                accelerated to higher energies ($\gamma \sim 10^5 -10^6$) 
                before they would upscatter X-ray photons emitted from the
                stellar surface.
                }
            \label{le_gamma}
        \end{figure}

        As we mentioned above, PSG can exist if equation 
        (\ref{heating_condition}) is satisfied. On the other hand for the
        heating of the stellar surface the high enough flux of back-streaming 
        particles is required. Thus, we need to estimate shielding factor 
        $\eta$ and gap height $h$ which are the main parameters of PSG. Let us
        note that $\eta$ and $h$ are connected with each other by equation 
        (\ref{gap_height_eqs}). Generally gap height can by defined as 
        $h \approx l_e + l_{ph}$ with the necessary condition that 
        $l_e > l_{acc}$. Latter corresponds to demand that the particle should
        gain the energy that is required for photon emission by either ICS or 
        CR processes. The Fig. \ref{le_gamma} shows dependence of free paths
        on 
        the particles Lorentz factor $\gamma$ for some particular pulsar 
        parameters (dependence on pulsar 
        parameters will be discussed in section \ref{observational_section}).
        Let us note that this free paths do not depend on the gap height $h$
        (see equations (\ref{cr_free_path}), (\ref{ics_free_path}), 
        (\ref{photon_path})). As it follows from equation 
        (\ref{acceleration_eqs}) $l_{acc}$ also does not depend on a gap height
        because $h$ and $\eta$ are connected by equation 
        (\ref{gap_height_eqs}).

        Results presented in the Fig. \ref{le_gamma} do not allow us to 
        define the gap height unambiguously but we can already find that CR is
        the dominant pair creation process for PSG scenario. Although 
        $l_{e, ics} < l_{e, cr}$ for particles with 
        $\gamma \sim 10^3 - 10^4$ the gamma-photon production by ICS process
        is not effective because $l_{\rm acc} << l_{e, ics}$, i.e. 
        the particles will be accelerated to higher energies 
        ($\gamma \sim 10^5 -10^6$) before they would upscatter X-ray photons 
        emitted from the hot polar cap.

        As soon as we determine the dominant process, i.e. CR, which is
        responsible for gamma-photons emission we can estimate the gap height.
        Curvature emission by a particle is effective for Lorentz factors
        $\gamma \sim 10^5 -10^6$ (when $l_{e,cr} \leq l_{\rm acc}$).
        The reaction force although is not high enough to stop the acceleration
        by electric field. Equilibrium between acceleration and 
        deceleration (by reaction force) would be established when the CR 
        power would equal to ''electric power'' ($P_{cr} = P_{\rm acc}$, where
        $P_{\rm acc} =  4 \pi \eta v B_s (h-z) / c P$ is work done by electric
        field in unit of time). Using the characteristic Lorentz factors of 
        radiating particles we can find characteristic frequencies of CR 
        photons and thus, we can check whether the necessary condition 
        ($l_e + l_{\rm ph} \leq h$) for cascade formation is satisfied. 

\begin{figure}[!ht]
                \centering
                \includegraphics[width=7.5cm, height=6.cm]{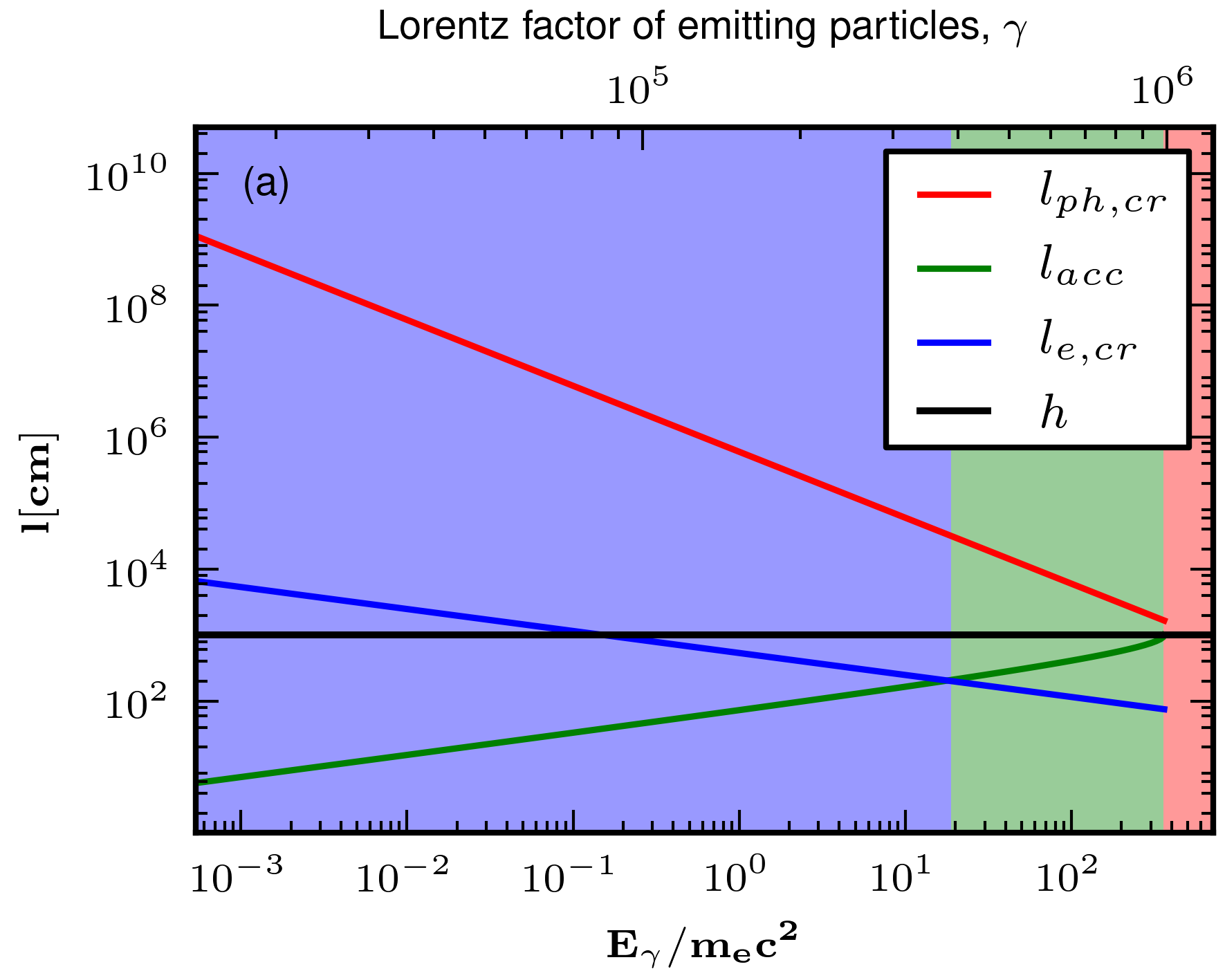}
                \includegraphics[width=7.5cm, height=6.cm]{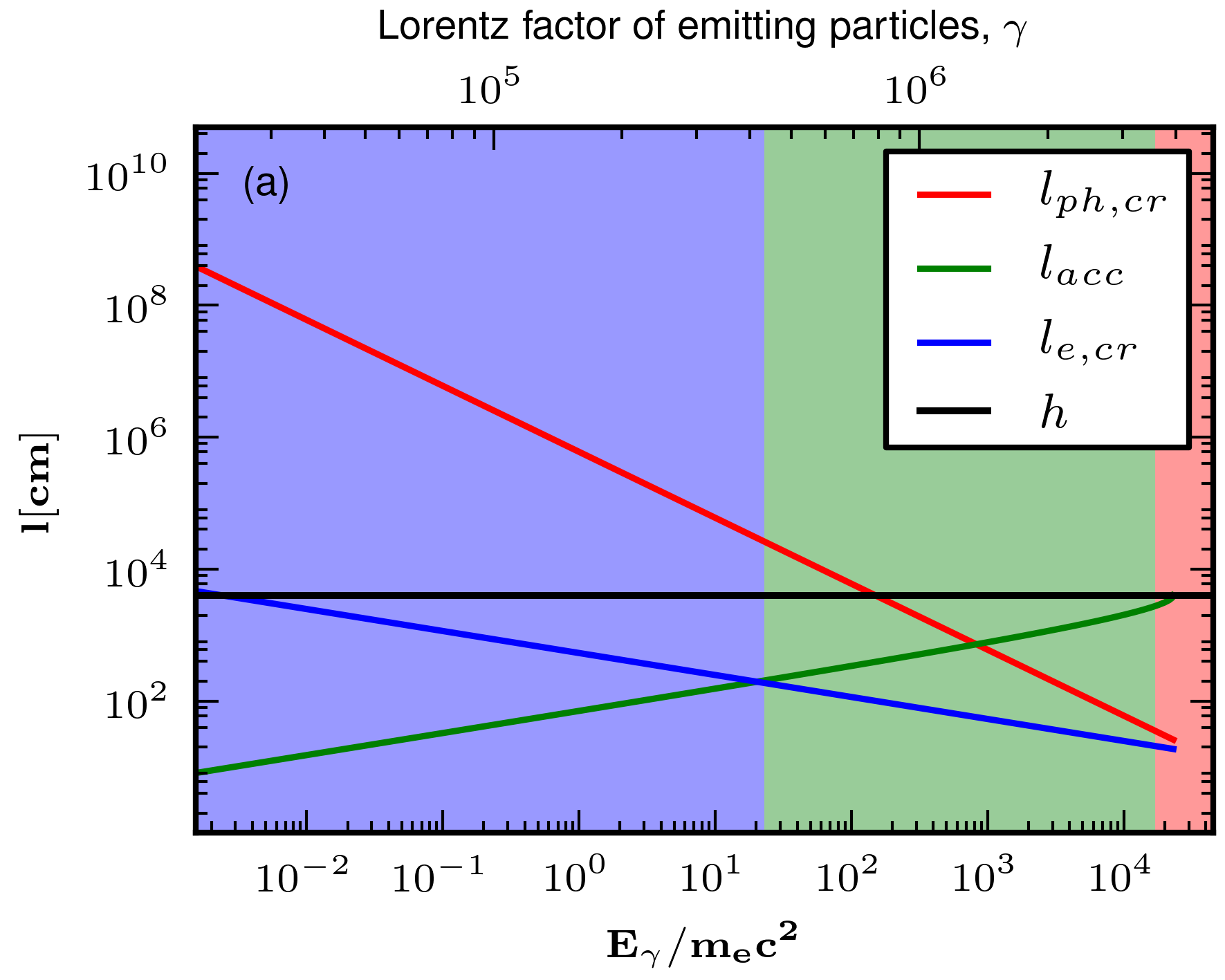}
                \caption{Diagram of the photon mean free path, the particle 
                    mean free path, and the acceleration path vs. the photon 
                    energy and the Lorentz factor of accelerated particles 
                    (top axis) for specific pulsar ($B_{14}=3.5$, $T_6=4.39$, 
                    $\Re_6=0.3$, $P=1$). See description of blue, green and 
                    red regions in the text above equation 
                    (\protect \ref{cascade_condition}). The black horizontal line
                    corresponds to the gap height. Calculations for panel (a) 
                    were done using $h=10$ m,  $\eta=0.03$ while for panel (b) 
                    using $h=40$ m, $\eta=0.01$.
                    } 
                \label{finding_height}
            \end{figure}

        The Fig. \ref{finding_height} shows dependence of the CR photon mean 
        free path ($l_{ph}$), the particle mean free path ($l_e$), and the 
        acceleration path ($l_{acc}$) on the energy expressed in units of 
        $m_e c^2$. Top axis
        shows the Lorentz factors of particles which emit the curvature photons
        with energy shown on bottom axis. While the energy of particle falls
        in blue region in the Fig. \ref{finding_height} there will be no 
        CR because the particle will be accelerated to higher energies before
        it travels distance enough to emit one curvature photon
        ($l_e>l_{\rm acc}$). For the energies in green region CR is most 
        efficient. The particle in the PSG never can reach energies falling in
        red region because reaction force caused by CR process is bigger then 
        acceleration force. Thus, the characteristic Lorentz factor $\gamma_c$
        of particles (and characteristic photon energy $E_{c}$) 
        corresponds to value defined by the border between green
        and red region. Therefore, we can define gap height for the 
        characteristic  energies $\gamma_c$ and $E_{c}$ as 
            \begin{equation}    
                l_{e}(\gamma=\gamma_c) + l_{ph}(E_{\gamma} = E_{c}) < h 
                \label{cascade_condition}
            \end{equation}
        Then we have to calculate $l_e$, $l_{ph}$ for different values of $h$ 
        and check which value satisfies condition (\ref{cascade_condition}).
        The Fig. \ref{finding_height} shows two cases: panel (a) that 
        corresponds
        to $h=10$ m and panel (b) which corresponds to  $h=40$ m. From panel 
        (a) we can see that
        condition (\ref{cascade_condition}) is not satisfied while for gap 
        height $h=40$ m cascade process will form.
        This approach allows us to calculate minimum height at which gap 
        sparking breakdown is possible and thus, we can estimate unambiguously
        the gap height in the frame of the PSG model.

    \section{Observational consequences \label{observational_section}}
        
        In order to compare the model with observations we need to define main
        parameters of PSG model for a given pulsar with known period and period
        derivative. 

        \subsection{PSG model parameters}
        
        We can distinguish two types of PSG parameters: observed and derived.
        As we mention above in some cases when X-ray observations are available
        we can directly estimate surface magnetic filed $B_s$. On one hand 
        $B_s$ can be calculated  using the size of the hot spot $A_{\rm BB}$, 
        and on the other hand we can find $B_s$ using estimation of the 
        critical temperature and assumption that $T_s = T_{\rm crit}$. One of
        the most important requirements of PSG model is that these two 
        estimations should coincide with each others. As it is clear from
        Fig. \ref{medin_lai} in most cases when the hot spot parameters are
        available this requirement is fulfilled. Thus, we can assume
        the characteristic values of $B_s$ vary in range of 
        $(1-4) \times 10^{14}$ which corresponds to critical/surface 
        temperature in range of $(1.1-4.5)\times 10^{6}$ 
        (see Table \ref{obs_table}). Using these values we can estimate derived
        parameters of PSG such us the gap height $h$, the shielding factor
        $\eta$ and the Lorentz factor of primary 
        particles $\gamma_c$. Let us note that these parameters depend also on 
        the curvature radius of magnetic field lines $\Re$. The curvature can
        not
        be neither observed or derived but modeling of surface magnetic field 
        (see Fig. \ref{non_dipolar_heat}) indicates that the curvature radius 
        varies in the range of $(0.1-1) \times 10^{6}$ cm. 
        Below we will discuss influence of pulsar parameters such
        as the magnetic field, the curvature of field lines and the period on
        derived PSG parameters.

            \subsection{Influence of the magnetic field}
\begin{figure}[!ht]
                \centering
                \includegraphics[height=4.9cm,width=4.95cm]{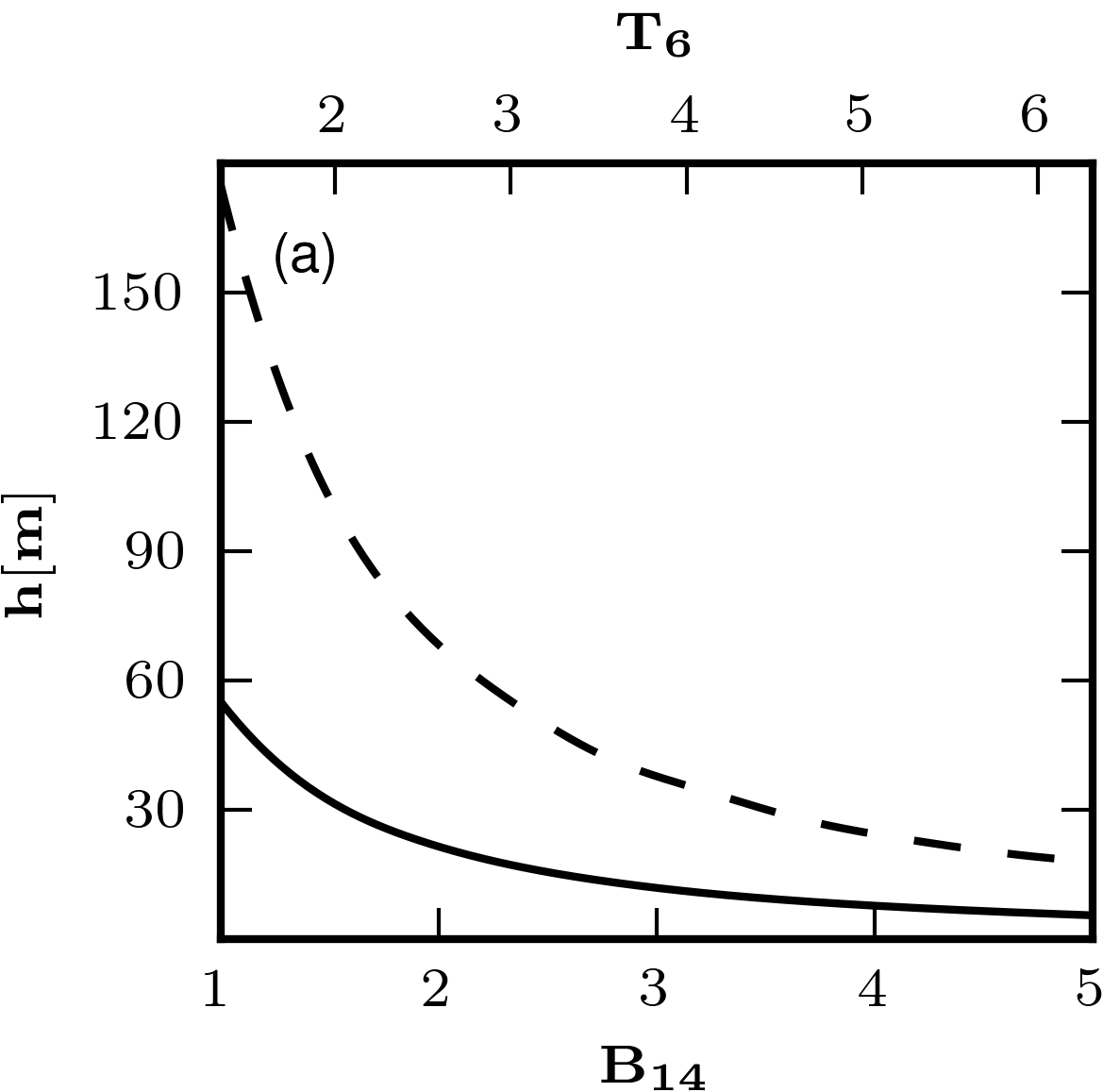}
                \includegraphics[height=4.9cm,width=4.95cm]{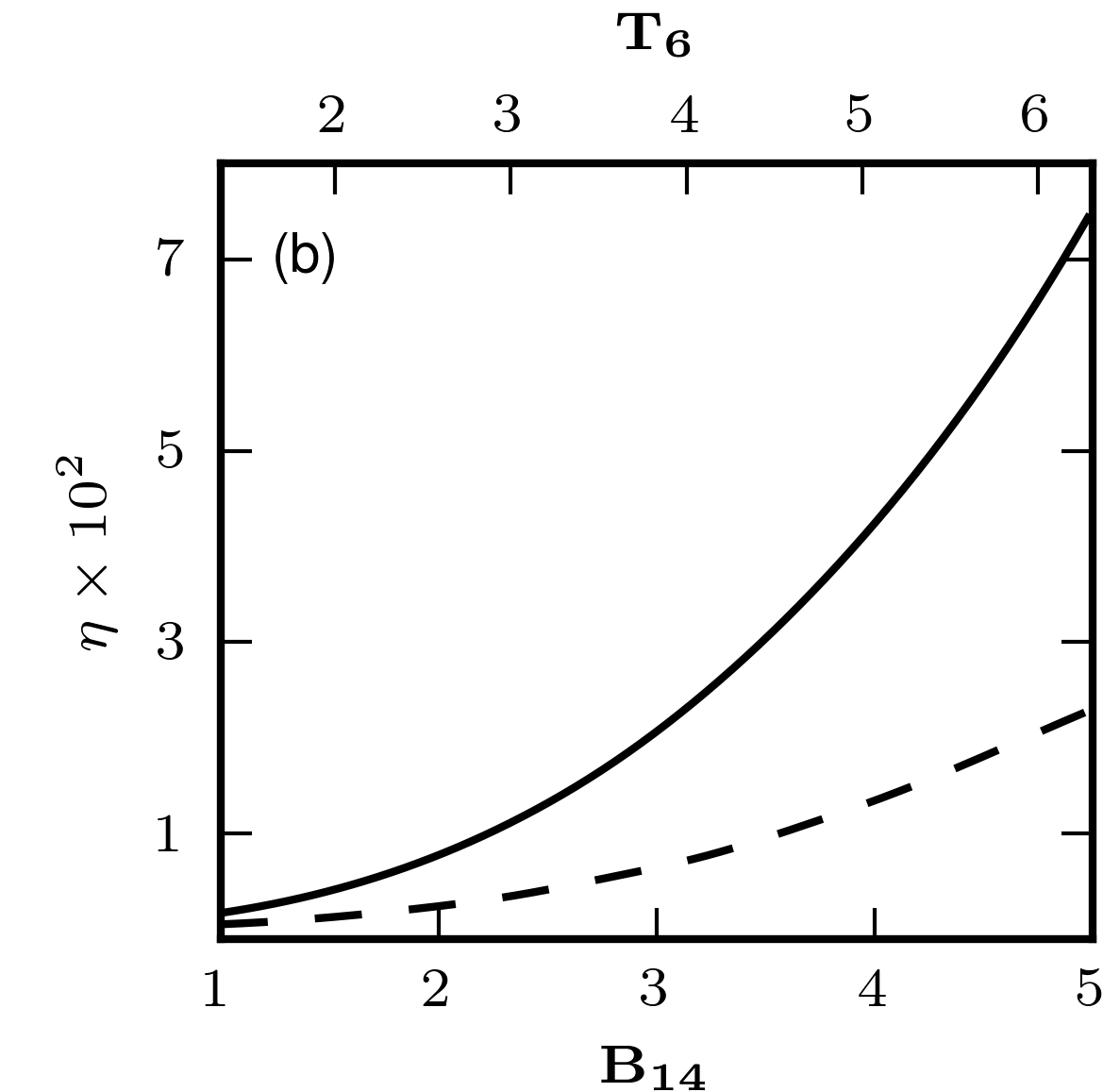}
                \includegraphics[height=4.9cm,width=4.95cm]{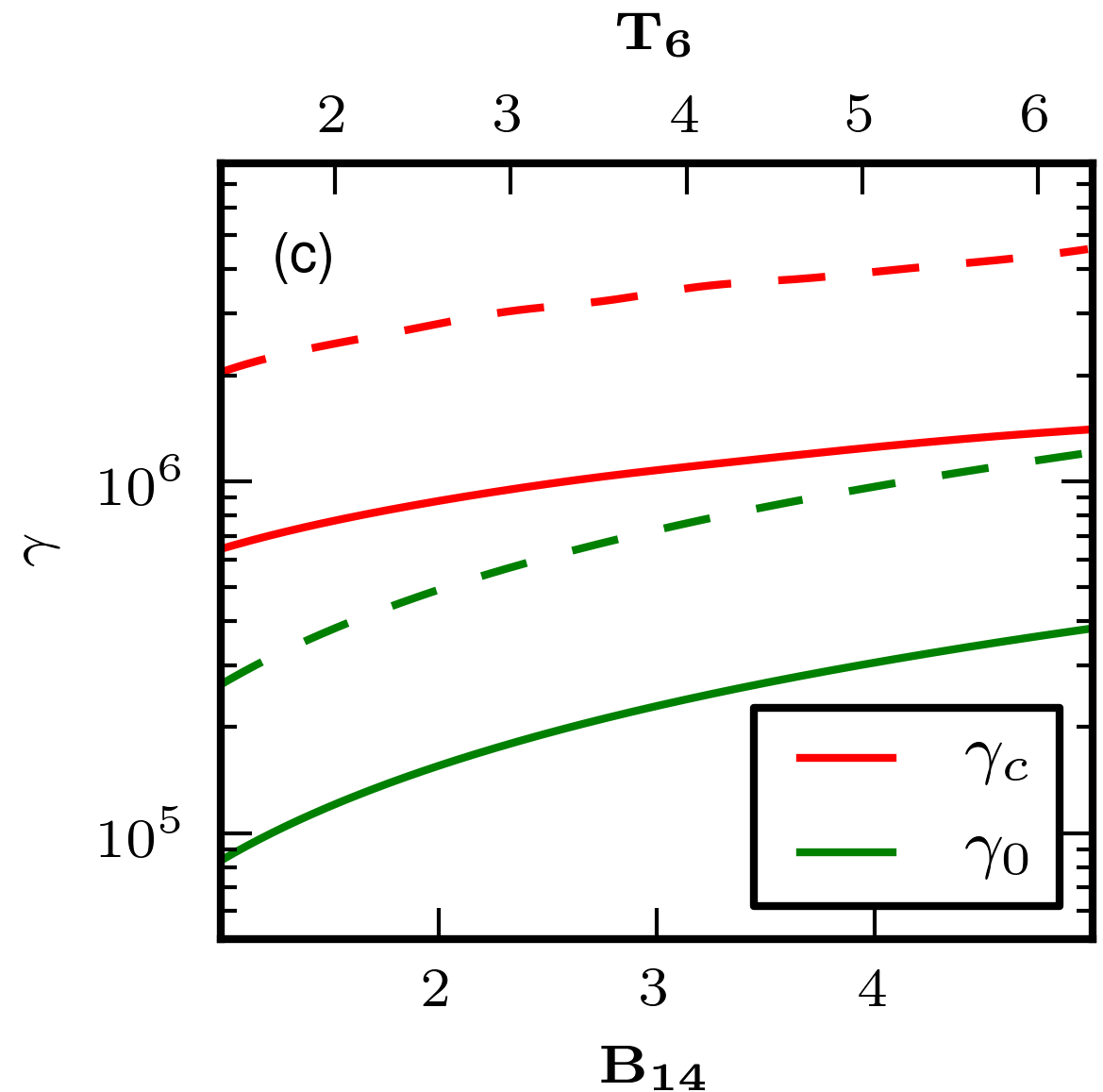}
                \caption{
                    Dependence of the gap height (panel a), the shielding 
                    factor (panel b), and the particles Lorentz factor 
                    (panel c) on the 
                    surface magnetic field. Solid lines correspond to 
                    calculations for $\Re_6=0.1$ while dashed lines
                    correspond to calculations for $\Re_6=1$. For both 
                    cases the pulsar period $P=1$ was used.
                    The green line presents the values corresponding to 
                    the boundary between blue and green region in the Fig. 
                    \protect \ref{finding_height} while the red line 
                    corresponds to the boundary between green and red region.
                    Corresponding critical temperature is shown on top axis of 
                    diagrams.
                    }
                    \label{h_eta_b14}
            \end{figure}

        The conditions in PSG are mainly defined by the surface magnetic field.
        In Fig. \ref{h_eta_b14} panel (a) we present dependence of the 
        gap height on the surface magnetic field calculated according to
        approach described in section \ref{finding_height_sec}. It is clear 
        that the gap height decreases as the surface magnetic field increases.
        The Fig. \ref{h_eta_b14} panel (b) shows dependence of the shielding
        factor on the surface magnetic field calculated using equation
        (\ref{gap_height_eqs}). We can see that for stronger magnetic fields
        $\eta$ increases which means that the density of heavy ions above the
        polar cap decreases. Let us note that the surface temperature ($T_s$) 
        stays very near to the critical temperature ($T_{\rm crit}$) which is 
        shown on the top axis of the diagrams.
        In Fig. \ref{h_eta_b14} panel (c) we present dependence of Lorenz
        factors of the particles accelerated in PSG. The Green line 
        ($\gamma_0$) presents the values corresponding to the boundary between
        blue and green region in the Fig. \ref{finding_height} while 
        the red line corresponds to the boundary between the green and the red
        region which is the characteristic value ($\gamma_c$) for primary
        particles. We see that $\gamma_c$ is very slightly affected by 
        the magnetic field strength.

        One can expect that for higher temperatures (which correspond to 
        stronger magnetic fields) the gap  breakdown could be dominated by ICS
        process. However this is not the case because the particle mean free
        path is much higher then the acceleration path 
        ($l_{e,ics} >> l_{\rm acc}$) even for strong magnetic fields.
        
            \subsection{Influence of the field lines curvature radius}
        
\begin{figure}[!ht]
                \centering
                \includegraphics[height=4cm,width=4.95cm]{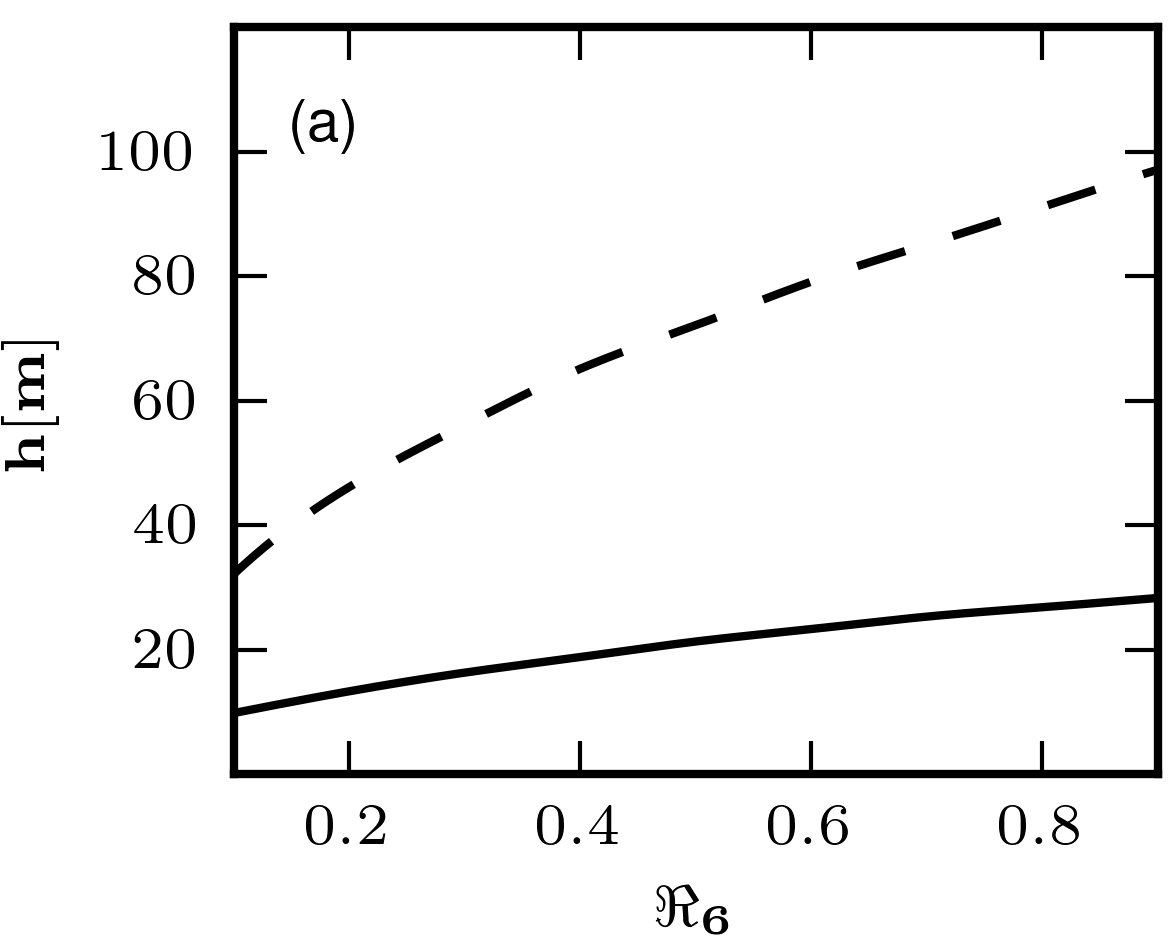}
                \includegraphics[height=4cm,width=4.95cm]{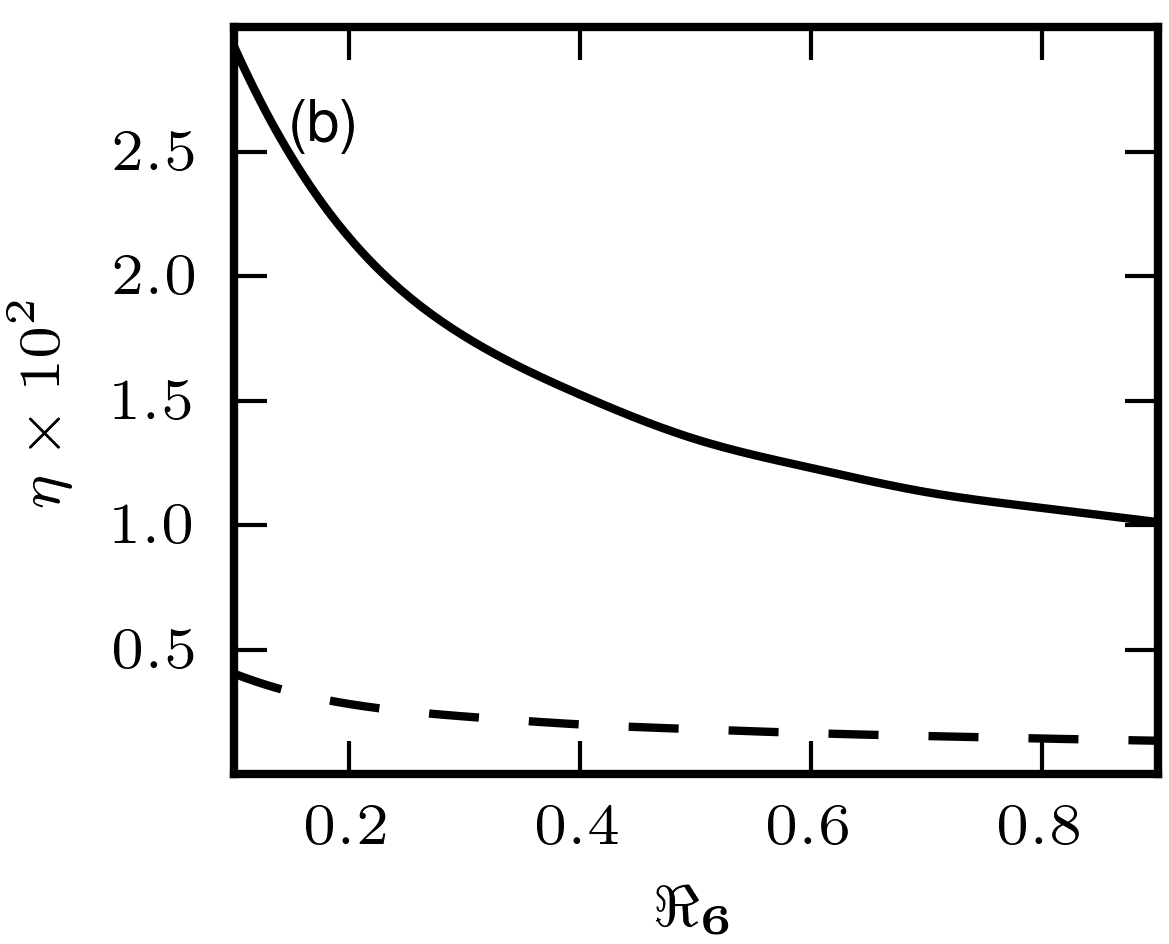}
                \includegraphics[height=4cm, width=4.95cm]{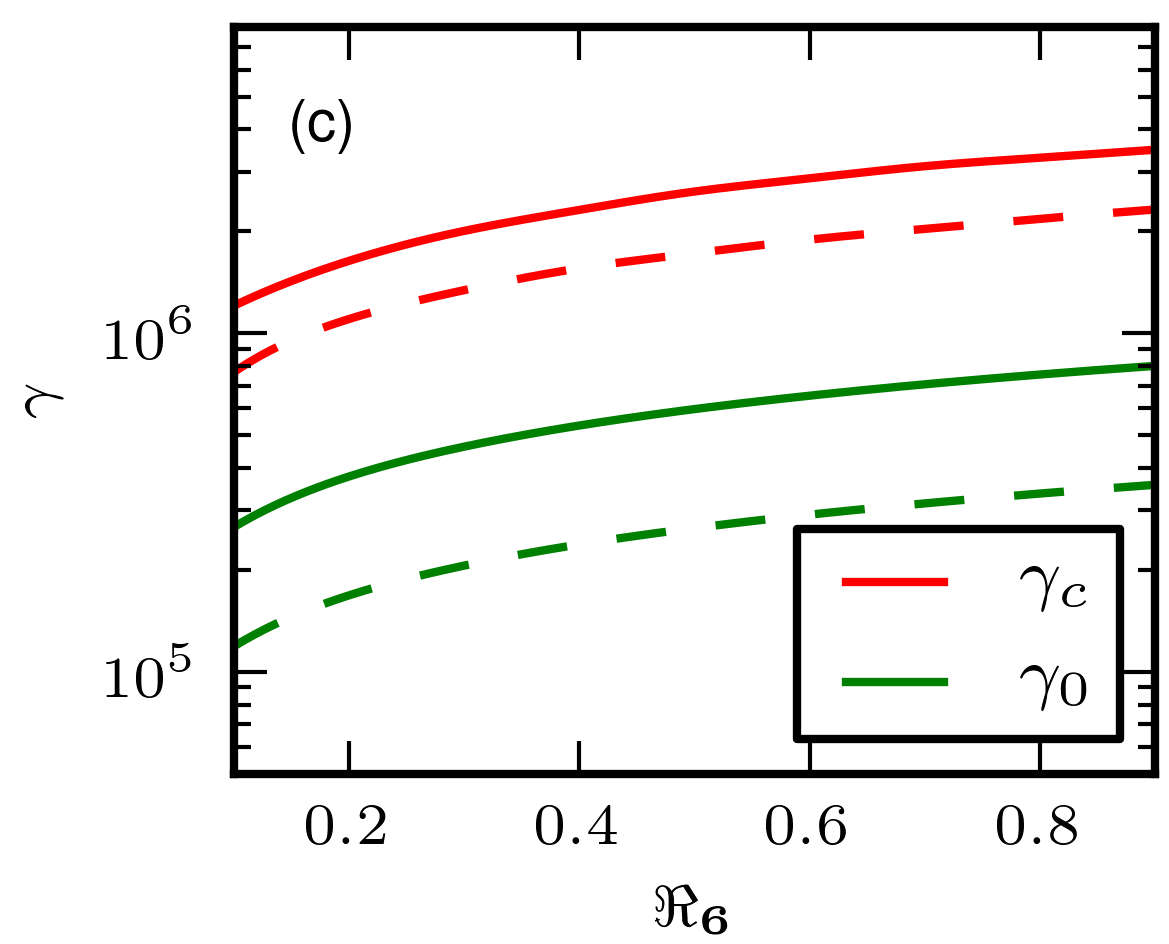}
                \caption{
                    Dependence of the gap height (panel a), the shielding 
                    factor (panel b), and the particles Lorentz factor 
                    (panel c) on the curvature radius of magnetic field lines.
                    Solid lines correspond to calculations for $B_{14}=3.5$ 
                    while dashed lines correspond to calculations for 
                    $B_{14}=1.5$. For both cases the pulsar period $P=1$ was 
                    used.
                    }
                \label{h_re6}
            \end{figure}

        The curvature of magnetic field lines significantly affects the
        PSG parameters since the CR process is responsible for PSG breakdown.
        As it can be expected the gap height decreases for smaller curvature 
        radius of magnetic field lines (Fig. \ref{h_re6} panel a) since CR is
        more efficient for smaller values of the curvature radius.
        Consequently the shielding factor  decreases with increasing curvature
        radius of magnetic field lines (Fig. \ref{h_re6} panel b).

        The Lorentz factor of primary particles increases for bigger radius of
        curvature (Fig. \ref{h_re6} panel c), since both conditions
        $l_{e} \sim l_{\rm acc}$ and $P_{cr} \sim P_{\rm acc}$ are satisfied
        for higher Lorentz factors. 
            
            \subsection{Influence of the pulsar period}

\begin{figure}[!ht]
                \centering
                \includegraphics[height=4cm, width=4.95cm]{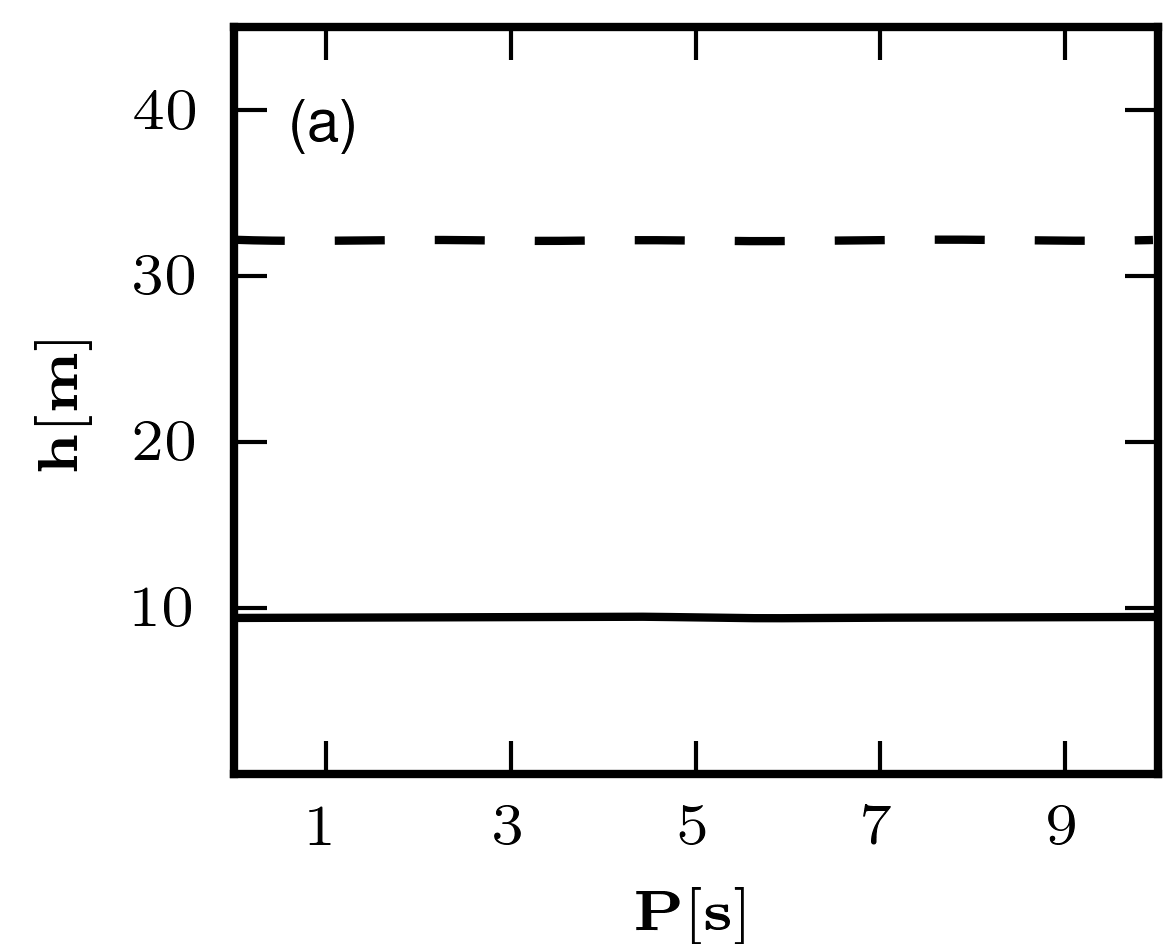}
                \includegraphics[height=4cm, width=4.95cm]{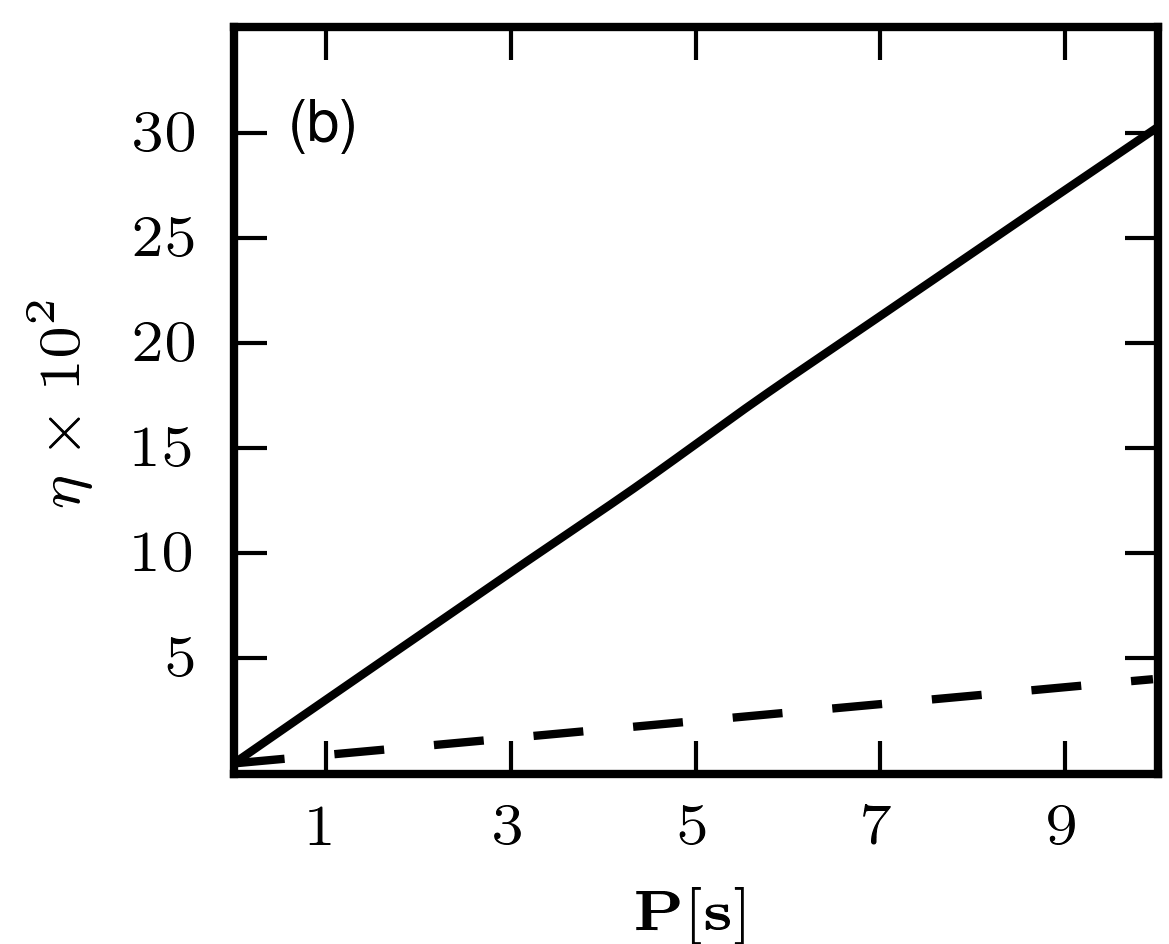}
                \includegraphics[height=4cm, width=4.95cm]{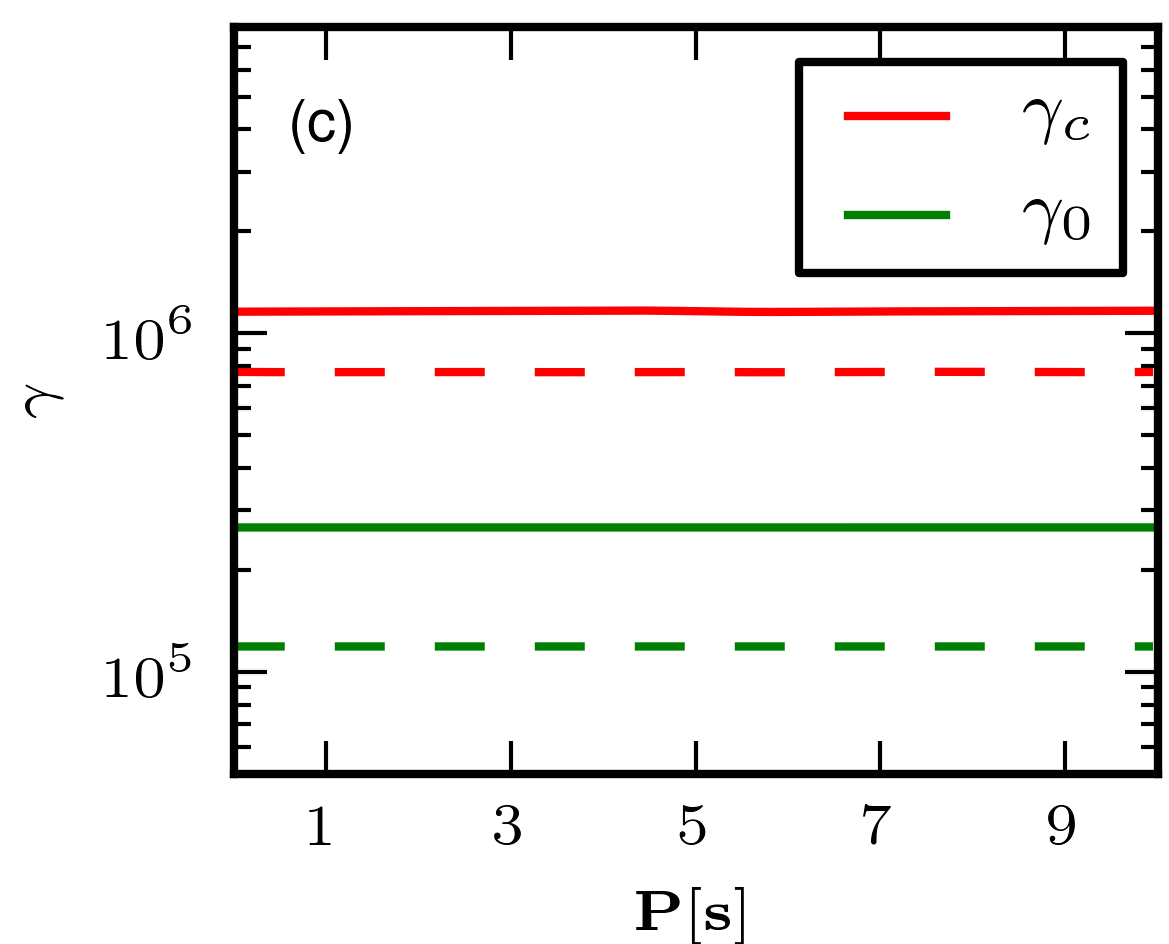}
                \caption{
                    Dependence of the gap height (panel a), the shielding 
                    factor (panel b), and the particles Lorentz factor 
                    (panel c) on the pulsar period.
                    Solid lines correspond to calculations for $B_{14}=3.5$
                    while dashed lines correspond to calculations 
                    for $B_{14}=1.5$. For both cases the radius of curvature 
                    $\Re_6=0.1$ was used.
                }
                \label{h_p}
            \end{figure}

        As we can see from figure \ref{h_p} panel (a) and panel (c) neither
        the gap height nor the Lorentz factor of primary particles depend on 
        pulsar period. This is one of the most important result of our 
        calculations. The shielding factor is just proportional to period 
        (Fig. \ref{h_p} panel b) as is expected from equation 
        (\ref{gap_height_eqs}).

    \section{Conclusions}
        
        X-ray observations of pulsars show that temperature of the hot spot
        is about few million kelvins while its area is much smaller then the 
        conventional polar cap surface which is calculated assuming purely 
        dipolar geometry of pulsar magnetic field.
        Such observations can be easily explained in the frame of PSG model 
        which assumes that the geometry of magnetic field near the stellar
        surface differs significantly from dipolar one, actually the field is 
        much stronger and curved. At the same time the surface temperature is
        about the critical temperature which depends only on the magnetic field
        strength.

        In this paper we calculated dependence of the gap parameters such as 
        the gap height, the shielding factor, and characteristic energies of 
        primary particles on the surface magnetic field, the curvature radius 
        of field lines, and the pulsar period in the frame of PSG model.
        The surface magnetic field can be calculated using the X-ray 
        observations (if such data are available), however the curvature radius
        can be only estimated by simulations of different geometry of magnetic
        field lines. Let us note that the gap height is the most important 
        parameter for all models of the Inner Acceleration Region in pulsars.
        We can define the gap height as a sum of the particle mean free path
        and the photon mean free path, $h \approx l_e + l_{ph}$.
        In order to estimate the PSG height we discuss two processes 
        responsible for cascade pair production: the Inverse Compton Scattering
        and the Curvature Radiation. We found that CR is the dominant process 
        for the sparking breakdown of the PSG. Although ICS is more efficient 
        for particles with $\gamma \sim 10^3 - 10^4$, the particles are 
        accelerated to higher energies $\gamma \sim 10^5 -10^6$ before they
        upscatter X-ray photons emitted from the polar cap. As soon as 
        particles Lorentz factor $\gamma > 10^5$ CR is more efficient then
        ICS. In section \ref{observational_section} we show the dependence of 
        PSG parameters ($h$, $\eta$, $\gamma_c$) on the pulsar parameters 
        ($B_s$, $\Re$, $P$). Since CR is the dominant process for gamma-photon
        emission the PSG parameters strongly depend on curvature radius of
        magnetic field lines. Pulsars with smaller curvature radius should have
        the lower gap height and also the higher shielding factor, consequently
        the density of heavy ions should be lower. In pulsars with stronger
        surface magnetic field gap heights should be smaller and also shielding
        factor should be smaller. Our calculations show that on one hand 
        $h$ and $\gamma_c$ do not depend on pulsar period, but on the other 
        hand $\eta$ increases along with increase pulsar period. 
        The evaluated PSG parameters should play decisive role in pulsar 
        emission models.

        We found that PSG model is suitable to explain both cases: when the hot
        spot is smaller then the conventional polar cap, and vice versa 
        when the hot spot is larger then conventional polar cap. In the latter
        case the surface is heated by particles created in the closed magnetic 
        filed lines region by photons emitted in the open magnetic filed lines
        region. Let us note that in the purely dipolar magnetic field geometry
        photons emitted tangent to magnetic field lines always stay in the 
        open field lines region.

    \acknowledgments{
        This paper was partially supported by research Grants N N 203 2738 33 
        and N N 203 3919 34 of the Polish Ministry of Science and Education. GM
        was partially supported by the GNSF grant ST08/4-442.
        }

    \begin{landscape}

        \begin{table}[t]
            \centering
            \small
            \begin{tabular}{lccccccccccc}
                name  &  $P$  &  $D$  &  $T_s$  &  $R_{\rm BB}$  &  $L_{\rm bol}$  &  $\chi$  &  $b$  &  $B_s$  &  $T_{\rm crit}$  &  age  &  ref.  \\
                &  $($ s $)$  &  $($ kpc $)$  &  $($ $10^6$ K $)$  &     &  $($ erg$/$s $)$  &   &   &  $($ $10^{14}$ G $)$  &  $($ $10^6$ K $)$  &  &   \\
                & & & & & & & & & & & \\
                \hline
                & & & & & & & & & & & \\
             
            J0108-1431  &  $0.808$  &  $0.18$  &  $3.2^{+0.4}_{-0.3}$  &  $6^{+5}_{-4} $ m  &  $7 \times 10^{27}$  &  $1 \times 10^{-3}$ &  $768$  &  $3.9$  &  $5.2$  &  $166 $ Myr  &  \cite{2009_Pavlov}
  \\
            B1929+10  &  $0.227$  &  $0.36$  &  $4.5^{+0.3}_{-0.5}$  &  $27^{+5}_{-4} $ m  &  $5 \times 10^{29}$  &  $1 \times 10^{-4}$ &  $129$  &  $1.3$  &  $1.8$  &  $3 $ Myr  &  \cite{2008_Misanovic}
  \\
            J0633+1746B  &  $0.237$  &  $0.16$  &  $2.3^{+0.1}_{-0.1}$  &  $36^{+9}_{-9} $ m  &  $8 \times 10^{28}$  &  $2 \times 10^{-6}$ &  $68$  &  $2.2$  &  $3.0$  &  $342 $ kyr  &  \cite{2005_Kargaltsev}
  \\
            & & & $0.5$ & $10$ km &  &  & & & & &  \\
            B0943+10  &  $1.098$  &  $0.63$  &  $3.1^{+1.1}_{-1.1}$  &  $18^{+40}_{-15} $ m  &  $5 \times 10^{28}$  &  $5 \times 10^{-4}$ &  $62$  &  $2.5$  &  $3.3$  &  $5 $ Myr  &  \cite{2006_Kargaltsev}
  \\
            B0950+08  &  $0.253$  &  $0.26$  &  $2.3^{+0.3}_{-0.3}$  &  $37^{+25}_{-25} $ m  &  $7 \times 10^{28}$  &  $1 \times 10^{-4}$ &  $59$  &  $0.3$  &  $0.6$  &  $18 $ Myr  &  \cite{2004_Zavlin}
  \\
            B1133+16  &  $1.188$  &  $0.36$  &  $3.2^{+0.5}_{-0.4}$  &  $18^{+14}_{-12} $ m  &  $7 \times 10^{28}$  &  $8 \times 10^{-4}$ &  $52$  &  $2.2$  &  $3.0$  &  $5 $ Myr  &  \cite{2006_Kargaltsev}
  \\
            B0834+06  &  $1.274$  &  $0.64$  &  $2.0^{+0.8}_{-0.6}$  &  $28^{+56}_{-15} $ m  &  $2 \times 10^{28}$  &  $2 \times 10^{-4}$ &  $21$  &  $1.2$  &  $1.7$  &  $3 $ Myr  &  \cite{2008_Gil}
  \\
            B0628-28  &  $1.244$  &  $1.45$  &  $3.3^{+1.3}_{-0.6}$  &  $59^{+65}_{-46} $ m  &  $7 \times 10^{29}$  &  $5 \times 10^{-3}$ &  $5$  &  $0.3$  &  $0.6$  &  $3 $ Myr  &  \cite{2005_Tepedelenl}
  \\
 & & & & & & & & & & & \\ 
 \hline 
 & & & & & & & & & & & \\ 

            J2043+2740  &  $0.096$  &  $1.80$  &  $1.5^{+0.4}_{-0.7}$  &  $467^{+200}_{-200} $ m  &  $2 \times 10^{30}$  &  $3 \times 10^{-5}$ &  $1$  &  --  &  --  &  $1 $ Myr  &  \cite{2004_Becker}
  \\
            & & & $0.6$ & $10$ km &  &  & & & & &  \\
            B1055-52  &  $0.197$  &  $0.75$  &  $1.8^{+0.1}_{-0.1}$  &  $460^{+60}_{-60} $ m  &  $4 \times 10^{30}$  &  $1 \times 10^{-4}$ &  $0.5$  &  --  &  --  &  $535 $ kyr  &  \cite{2005_Deluca}
  \\
            & & & $0.8$ & $12 $ km &  &  & & & & &  \\
            J0538+2817  &  $0.143$  &  $1.20$  &  $2.8^{+0.1}_{-0.1}$  &  $666^{+38}_{-38} $ m  &  $5 \times 10^{31}$  &  $9 \times 10^{-4}$ &  $0.3$  &  --  &  --  &  $30 $ kyr  &  \cite{2003_Mcgowan}
  \\
            J1809-1917  &  $0.083$  &  $3.50$  &  $2.0^{+0.4}_{-0.4}$  &  $951^{+920}_{-693} $ m  &  $2 \times 10^{31}$  &  $1 \times 10^{-5}$ &  $0.3$  &  --  &  --  &  $51 $ kyr  &  \cite{2007_Kargaltsev}
  \\
            J0821-4300  &  $0.113$  &  $2.20$  &  $6.3^{+0.2}_{-0.2}$  &  $1.2^{+0.1}_{-0.1} $ km  &  $4 \times 10^{33}$  &  $1 \times 10^{-1}$ &  $0.1$  &  --  &  --  &  $2 $ Myr  &  \cite{2010_Gotthelf}
  \\
            & & & $3.2$ & $6.0 $ km &  &  & & & & &  \\
            B0833-45  &  $0.089$  &  $0.30$  &  $1.9^{+0.1}_{-0.1}$  &  $2.0^{+0.2}_{-0.2} $ km  &  $3 \times 10^{31}$  &  $5 \times 10^{-6}$ &  $0.06$  &  --  &  --  &  $11 $ kyr  &  \cite{2007_Zavlin_b}
  \\
            & & & $0.9$ & $14 $ km &  &  & & & & &  \\
            J1357-6429  &  $0.166$  &  $2.50$  &  $2.2^{+0.3}_{-0.3}$  &  $1.9^{+0.4}_{-0.4} $ km  &  $2 \times 10^{32}$  &  $5 \times 10^{-5}$ &  $0.03$  &  --  &  --  &  $7 $ kyr  &  \cite{2007_Zavlin}
  \\
            B1916+14  &  $1.181$  &  $2.10$  &  $1.5^{+0.1}_{-0.1}$  &  $800^{+100}_{-100} $ m  &  $6 \times 10^{30}$  &  $1 \times 10^{-3}$ &  $0.03$  &  --  &  --  &  $88 $ kyr  &  \cite{2009_Zhu}
  \\
            B1706-44  &  $0.102$  &  $2.50$  &  $2.2^{+0.2}_{-0.2}$  &  $2.8^{+0.7}_{-0.7} $ km  &  $3 \times 10^{32}$  &  $9 \times 10^{-5}$ &  $0.03$  &  --  &  --  &  $18 $ kyr  &  \cite{2002_Gotthelf}
  \\
            B0656+14  &  $0.385$  &  $0.29$  &  $1.2^{+0.03}_{-0.03}$  &  $1.8^{+0.2}_{-0.2} $ km  &  $1 \times 10^{31}$  &  $4 \times 10^{-4}$ &  $0.02$  &  --  &  --  &  $111 $ kyr  &  \cite{2005_Deluca}
  \\
            & & & $0.7$ & $21 $ km &  &  & & & & &  \\
            B2334+61  &  $0.495$  &  $3.10$  &  $1.7^{+0.5}_{-0.8}$  &  $1.7^{+0.6}_{-0.4} $ km  &  $1 \times 10^{31}$  &  $2 \times 10^{-4}$ &  $0.02$  &  --  &  --  &  $41 $ kyr  &  \cite{2006_Mcgowan}
  \\
            J1119-6127  &  $0.408$  &  $8.40$  &  $3.1^{+0.4}_{-0.3}$  &  $2.6^{+1.4}_{-0.2} $ km  &  $1 \times 10^{33}$  &  $5 \times 10^{-4}$ &  $0.008$  &  --  &  --  &  $2 $ kyr  &  \cite{2007_Gonzalez}
  \\
            \end{tabular}
            \caption{\small{
                Observed and derived parameters of isolated neutron stars with
                available X-ray observations. $P$ - the pulsar period, $D$ -
                the distance used for calculation of $L_{\rm bol}$, $T_s$ - the
                spot temperature, $R_{\rm BB}$ - the radius of the spot
                obtained from the black-body fit,  $L_{\rm bol}$ - the
                bolometric luminosity of the hot or warm spot
                ($L_{\rm bol} = \pi R_{\rm BB}^2\sigma T_s^4 $),
                $\chi = L_{bol} / L_{sd}$ - the
                efficiency of X-ray emission where $L_{sd}$ is the spin down
                luminosity, $b = A_{\rm pc} / A_{\rm BB} = B_s / B_d$, $B_s$ -
                the surface magnetic field strength, $T_{\rm crit}$ - the
                critical temperature. Pulsars are sorted by decreasing
                parameter $b$.}
            }
            \label{obs_table}
        \end{table}

    \end{landscape}

\end{document}